\documentclass[eqsecnum,showpacs,amsmath,aps,nofootinbib]{revtex4}  
\usepackage{epsfig}
\usepackage{amssymb}
\usepackage[dvips]{color}
\usepackage[latin1]{inputenc}
\usepackage{graphicx}
\usepackage{gensymb}
\usepackage{amsmath}
\usepackage{subfig}
\usepackage{relsize}
 \usepackage{mathtools}

\def\beq{\begin{equation}}
\def\eeq{\end{equation}}
\def\beqq{\begin{eqnarray}}
\def\eeqq{\end{eqnarray}}
\def\D{\partial}
\def\A{\mathcal{A}_{ir}}
\def \H{\mathcal{H}_i}
\def \dd {\dfrac{{\rm d}}{{\rm d}t}}
\def \DD {\dfrac{{\rm d}^2}{{\rm d}t^2}}

\newcommand{\bdm}{\begin{displaymath}}
\newcommand{\edm}{\end{displaymath}}

\def\pmb#1{\setbox0=\hbox{$#1$}%
  \kern-.025em\copy0\kern-\wd0
  \kern.05em\copy0\kern-\wd0
  \kern-.025em\raise.0433em\box0}

\def\D{\partial}

\makeatletter
\renewcommand*{\@fnsymbol}[1]{\ensuremath{\ifcase#1\or *\or \dagger\or
    \ddagger\or 
   \mathsection\or **\or \dagger\dagger
   \or \ddagger\ddagger \else\@ctrerr\fi}}
\makeatother

\begin{document}
\title{Solar system dynamics in general relativity} 

\author{Emmanuele Battista}
\email[E-mail: ]{ebattista@na.infn.it}
\affiliation{Istituto Nazionale di Fisica Nucleare, Sezione di Napoli, Complesso Universitario di Monte
S. Angelo, Via Cintia Edificio 6, 80126 Napoli, Italy}

\author{Simone Dell'Agnello}
\email[E-mail: ]{simone.dellagnello@lnf.infn.it}
\affiliation{Istituto Nazionale di Fisica Nucleare, Laboratori Nazionali di Frascati,
00044 Frascati, Italy}

\author{Giampiero Esposito}
\email[E-mail: ]{gesposit@na.infn.it}
\affiliation{Istituto Nazionale di Fisica Nucleare, Sezione di
Napoli, Complesso Universitario di Monte S. Angelo, 
Via Cintia Edificio 6, 80126 Napoli, Italy}

\author{Jules Simo}
\email[E-mail: ]{JSimo@uclan.ac.uk}
\affiliation{Aerospace Engineering, Computing \& Technology Building,\\
School of Engineering, University of Central Lancashire, Preston, PR1 2HE, United Kingdom}

\author{Luciano Di Fiore}
\email[E-mail: ]{luciano.difiore@na.infn.it}
\affiliation{Istituto Nazionale di Fisica Nucleare, Sezione di Napoli,
Complesso Universitario di Monte S. Angelo,
Via Cintia Edificio 6, 80126 Napoli, Italy}

\author{Aniello Grado}
\email[E-mail: ]{aniello.grado@gmail.com}
\affiliation{INAF, Osservatorio Astronomico di Capodimonte, 80131 Napoli, Italy}

\date{\today}

\begin{abstract}
Recent work in the literature has advocated using the Earth-Moon-planetoid
Lagrangian points as observables, in order to test general relativity and
effective field theories of gravity in the solar system. However, since the three-body problem 
of classical celestial mechanics is just an approximation of a much
more complicated setting, where all celestial bodies in the solar system 
are subject to their mutual gravitational interactions, 
while solar radiation pressure and other sources of nongravitational perturbations 
also affect the dynamics, it is conceptually desirable to improve the current understanding of solar
system dynamics in general relativity, as a first step towards a more accurate theoretical study of
orbital motion in the weak-gravity regime. For this purpose, starting from the Einstein equations in
the de Donder-Lanczos gauge, this paper arrives first at the Levi-Civita Lagrangian for the geodesic
motion of celestial bodies, showing in detail under which conditions the effects of internal structure 
and finite extension get cancelled in general relativity to first post-Newtonian order.
The resulting nonlinear ordinary differential equations for the motion of planets and satellites 
are solved for the Earth's orbit about the Sun, written down in detail for the Sun-Earth-Moon system,
and investigated for the case of planar motion of a body immersed in the gravitational field produced
by the other bodies (e.g. planets with their satellites). At this stage, we prove an exact property, 
according to which the fourth-order time derivative of the original system leads to a linear system of
ordinary differential equations. This opens an interesting perspective on forthcoming research on
planetary motions in general relativity within the solar system, although the resulting equations 
remain a challenge for numerical and qualitative studies. Last, the evaluation of quantum corrections 
to location of collinear and noncollinear Lagrangian points for the planar restricted three-body
problem is revisited, and a new set of theoretical values of such corrections for the Earth-Moon-planetoid
system is displayed and discussed. On the side of classical values, the general relativity corrections
to Newtonian values for collinear and noncollinear Lagrangian points of the Sun-Earth-planetoid system
are also obtained.
\end{abstract}


\maketitle

\section{Introduction}

Recent work by the authors \cite{B1,B2,B3,B4}, motivated, on the quantum side, by modern developments
in effective field theories of gravity \cite{EG1,EG2,EG3,EG4,EG5,EG6,EG7,EG8,EG9}, and, on the classical
side, by the beautiful discoveries in celestial mechanics 
\cite{CM1,CM2,CM3,CM4,CM5,CM6,CM7,CM8,CM9,CM10,CM11,CM12,CM13,CM14,CM15,CM16,CM17,CM18,CM19,CM20,
CM21,CM22,CM23,CM24,CM25,CM26,CM27,CM28,CM29,CM30,CM31,CM32,CM33,CM34,CM35,CM36,CM37,CM38,CM39,CM40,CM41},
aerospace engineering \cite{OR1,OR2,OR3,OR4,OR5,OR6,OR7,OR8} and (lunar) laser ranging techniques
\cite{LR1,LR2,LR3,LR4,LR5,LR6,LR7,LR8,LR9,LR10,LR11,LR12,LR13}, has studied in detail the libration
points of the restricted three-body problem in Newtonian gravity, general relativity and effective
field theories of gravity. In particular, we have found 
(see Ref. \cite{B4} and Erratum therein) that general relativity corrects
by $0.19$ mm, $-0.32$ mm and $-0.04$ mm the location of the collinear Lagrangian points $L_{1},L_{2}$ 
and $L_{3}$ respectively, for the circular restricted three-body problem where a planetoid (e.g. a
satellite) moves in the gravitational field of the Earth and the Moon. 
Moreover, for the planar $(x,y)$ coordinates
of noncollinear Lagrangian points $L_{4}$ and $L_{5}$, the Einstein correction to Newtonian values is
$2.73$ mm for $x$ and $-1.59$ mm for $y$. The possible quantum corrections\footnote{These might be
seen as low-energy effects in quantum gravity, more easily accessible than the elusive high-energy
effects of the Planck era.} to the Lagrangian points $L_{1}...L_{5}$ are just below or just above 
$1$ mm. These tiny theoretical values are conceptually interesting but, unfortunately, too small to put
on firm ground the hope that one might arrive in the near future to a new test of general relativity or to
a clearcut discrimination between corrections predicted by general relativity and those predicted by
effective field theories of gravity.

Nevertheless, since the three-body problem is just an approximation of a much more complicated setting, where 
all celestial bodies in the solar system (planets, satellites, asteroids, ...) interact with 
each other while solar radiation pressure \cite{SUN} 
and other sources of nongravitational perturbations may
also come into play, it remains conceptually desirable to improve the current understanding of solar 
system dynamics in general relativity, as a first step towards a more accurate theoretical study
of orbital motion, possibly including the outstanding open problem of solar system stability.  

For this purpose, Sec. II performs a concise but detailed review of the $N$-body problem in general
relativity, following the thorough analysis in Ref. \cite{CM6}. First, the Einstein equations are
studied in the de Donder-Lanczos gauge so as 
to obtain a Lagrangian for the geodesic motion of celestial bodies. Second, the physics of
gravitating bodies with finite extension is studied, showing in detail under which conditions
one arrives at a cancellation effect also in classical general relativity.
Section III applies the Lagrangian of Sec. II to the Earth's motion around the Sun and then
to the Sun--Earth-Moon system, and eventually to the planar motion of a body subject to the
gravitational attraction of several bodies, when all their mutual distances are large enough
that we are in the weak-gravity regime. Section IV derives in detail the linear 
(not linearized!) system of ordinary differential equations associated with our nonlinear 
equations of motion. Section V considers another possible definition
of quantum corrections to Lagrangian points. Results and open problems are discussed in Sec. VI.  

\section{Lagrangian of the N-body problem in general relativity}

In General Relativity, the N-body Lagrangian of celestial mechanics is derived by appealing
to the geodesic principle for the motion of each celestial body. The first part of the analysis
deals with the field equations and derives an approximate form of the metric components and of
the resulting line element under suitable assumptions. The second part of the analysis relies upon
three basic assumptions on the physics of gravitating systems, which make it precise what sort of
cancellation principle still holds on passing from Newtonian to relativistic celestial mechanics.
Hence a Lagrangian of $N$ material points is eventually derived.

\subsection{The Einstein equations}

The Einstein equations
\begin{equation}
R_{\mu \nu}-{1 \over 2}g_{\mu \nu}R={8 \pi G_{N}\over c^{4}}T_{\mu \nu}
\label{(2.1)}
\end{equation}
are a quasi-linear set of partial differential equations. As is well known, this means that they are
linear in second-order derivatives of the metric $g_{\mu \nu}$, whereas the nonlinearity results
from the squares of first derivatives of the spacetime metric. Such a quasi-linear system is not in
normal form unless a suitable supplementary condition (more frequently called gauge-fixing) is imposed.
In the Levi-Civita analysis we rely upon, such a condition is the so-called de Donder-Lanczos gauge. 
The Lanczos gauge sets to zero the action of the scalar wave operator $\Box$ on spacetime
coordinates, i.e.
\begin{equation}
\Box x^{\nu}=-\sum_{\lambda,\mu=0}^{3}g^{\lambda \mu}\Gamma_{\; \lambda \mu}^{\nu}=0.
\label{(2.2)}
\end{equation}
Such a set of spacetime coordinates, if they exist, were said to be isometric because Eqs. (2.2)
generalize the harmonic character of Cartesian coordinates in a Euclidean metric. The de Donder formulation
takes linear combinations of such equations with coefficients given by covariant components of 
the metric, i.e.
\begin{equation}
Z_{\mu}=\sum_{\nu=0}^{3}g_{\mu \nu}\Box x^{\nu}.
\label{(2.3)}
\end{equation}
Now a remarkable identity relates the Ricci tensor and the symmetrized 
partial derivatives of the $Z_{\mu}$, i.e.
\begin{equation}
R_{\mu \nu}-{1 \over 2}\left({\partial Z_{\mu}\over \partial x^{\nu}}
+{\partial Z_{\nu}\over \partial x^{\mu}}\right)
={1 \over 2}\sum_{\lambda,\rho=0}^{3}g^{\lambda \rho}{\partial^{2}g_{\mu \nu}
\over \partial x^{\lambda} \partial x^{\rho}}
+F\left(g_{\alpha \beta},{\partial g_{\mu \nu}\over \partial x^{\rho}}\right).
\label{(2.4)}
\end{equation}
The term $F$ is of lower order in that it depends only on the metric and its first partial
derivatives. In the desired isometric coordinates, for which the $Z_{\mu}$ vanish, the Einstein 
equations, jointly with the conservation equations for the energy-momentum tensor
resulting from the Bianchi identity, take therefore the normal
form \cite{CM6} with respect to the $x^{0}$ variable.

In the following analysis, the dimensionless terms with an order of magnitude of
\begin{equation}
\beta^{2} \equiv {v^{2}\over c^{2}}, \;
\gamma \equiv {G_{N}\over c^{2}}\int_{S}{\mu \over r}{\rm d}S 
={G_{N}\over c^{2}}
\sum_{h=1}^{n}\int_{h}{\mu \over r}{\rm d}C_{h},
\label{(2.5)}
\end{equation}
where $S$ is the region occupied by all bodies $C_{h}$, $h=1,...,n$, and $\mu$ is the function
representing\footnote{No confusion should arise when we write instead 
$\mu$ for a coordinate index in tensor equations, and similarly for other Greek letters used
hereafter, whose meaning will be clear from the context.}
the local density, are said to be of first order
and are denoted by ${\rm O}(I)$. The use of the de Donder-Lanczos gauge implies that the
metric components take the approximate form \cite{CM6}
\begin{equation}
g_{00}=1-2 \gamma +2 \gamma^{2}-2 \zeta,
\label{(2.6)}
\end{equation}
\begin{equation}
g_{0i}=g_{i0}=4 \gamma_{i}+{\rm O}(II),
\; \forall i=1,2,3,
\label{(2.7)}
\end{equation}
\begin{equation}
g_{ii}=-1-2 \gamma, \; \forall i=1,2,3.
\label{(2.8)}
\end{equation}
With this notation, $\gamma$ is the Newtonian potential in a domain $S$ of Euclidean 
space\footnote{It is a nontrivial property that three-dimensional space is taken to be
Euclidean. From the point of view of general formalism, some authors regard this assumption
as Procuste's bed \cite{CM13} and hence a severe drawback, but for the analysis of planetary
motions this remains a legitimate approach, whereas it would be totally inappropriate in the
relativistic astrophysics of binary systems.}
with Cartesian coordinates $x^{1},x^{2},x^{3}$, and ${\rm O}(II)$ is a term of
second order in $\beta^{2}$ and $\gamma$. 
If $x^{0} \equiv ct$, the associated dimensionless velocities are
\begin{equation}
\beta^{i} \equiv {{\rm d}x^{i}\over {\rm d}x^{0}}
\Longrightarrow \beta^{2} \equiv \sum_{i=1}^{3}(\beta^{i})^{2}.
\label{(2.9)}
\end{equation}
One can therefore define the three potentials
\begin{equation}
\gamma_{i} \equiv {G_{N}\over c^{2}} \int_{S} {\mu \beta_{i}\over r}{\rm d}S,
\label{(2.10)}
\end{equation}
while the function $\zeta$ can be split into the sum of three functions,
i.e. \cite{CM6}
\begin{equation}
\zeta \equiv \varphi+\psi + \nu,
\label{(2.11)}
\end{equation}
where $\varphi$ and $\psi$ solve a Poisson-type equation:
\begin{equation}
\varphi \equiv -{G_{N}\over c^{2}} \int_{S}{\mu \gamma \over r}{\rm d}S,
\label{(2.12)}
\end{equation}
\begin{equation}
\psi \equiv {3 \over 2}{G_{N}\over c^{2}}
\int_{S}{\mu \beta^{2}\over r}{\rm d}S,
\label{(2.13)}
\end{equation}
whereas
\begin{equation}
\nu \equiv {1 \over 2}{G_{N}\over c^{2}}
{\partial^{2}\over \partial (x^{0})^{2}}\int_{S}\mu r {\rm d}S.
\label{(2.14)}
\end{equation}
By virtue of (2.6)-(2.14), the squared line element takes the approximate form
\begin{equation}
\left({{\rm d}s \over {\rm d}x^{0}}\right)^{2}
=1-2 \left({\beta^{2}\over 2}+\gamma \right)+2 \gamma^{2}-2 \zeta -2 \gamma \beta^{2}
+8 \sum_{i=1}^{3}\gamma_{i}\beta^{i}.
\label{(2.15)}
\end{equation}
The formulation of the geodesic principle for celestial bodies needs actually the square root of (2.15),
i.e. 
\begin{equation}
\delta \int {\rm d}s=\delta \int {{\rm d}s \over {\rm d}x^{0}}{\rm d}x^{0}=0.
\label{(2.16)}
\end{equation}
From the standard second-order Taylor expansion of $(1+\varepsilon)^{1 \over 2}$
in the neighboorhood of $\varepsilon=0$ one gets, upon denoting by ${\cal N}$ the
Newtonian gravitational Lagrangian
\begin{equation}
{\cal N} \equiv {\beta^{2}\over 2}+\gamma,
\label{(2.17)}
\end{equation}
the useful approximate formula
\begin{eqnarray}
{{\rm d}s \over {\rm d}x^{0}}&=&
\sqrt{1-2 \left({\beta^{2}\over 2}+\gamma \right)+2 \gamma^{2}-2 \zeta -2 \gamma \beta^{2}
+8 \sum_{i=1}^{3}\gamma_{i}\beta^{i}}
\nonumber \\
& \approx & 1-{\cal N}+\gamma^{2}-\zeta -\gamma \beta^{2}
+4 \sum_{i=1}^{3}\gamma_{i}\beta^{i}-{1 \over 2}{\cal N}^{2}.
\label{(2.18)}
\end{eqnarray}
The approximate calculation leading from (2.15) to (2.18) is simple, but it represents a
crucial conceptual step. A rigorous analysis of stability of the solar system would require
working with the square root on the first line of Eq. (2.18) without any expansion, and then
using the modern qualitative methods of the calculus of variations.
The constant $1$ is of course inessential, and this is made precise be pointing out that
also the following variation vanishes \cite{ADC}:
\begin{equation}
\delta \int {\rm d}x^{0}=0,
\label{(2.19)}
\end{equation}
because we can avoid letting $x^{0}$ to vary since the left-hand side of the equation expressing
the geodesic principle undergoes a variation that vanishes by virtue of the conditions resulting
from variation of the three coordinates $x^{1},x^{2},x^{3}$. In light of (2.16) and (2.19),
we get eventually the dimensionless Lagrangian
\begin{equation}
{\cal L}=1-{{\rm d}s \over {\rm d}x^{0}}={\cal N}+{\cal D},
\label{(2.20)}
\end{equation}
where ${\cal N}$, defined in (2.17), is the Lagrangian of a dimensionless material element in 
Newtonian mechanics, whereas $\cal D$ is the Einstein modification of this Newtonian
Lagrangian, and is given by \cite{CM6}
\begin{equation}
{\cal D} \equiv {1 \over 2} {\cal N}^{2}-\gamma^{2}+\zeta+\gamma \beta^{2}
-4 \sum_{i=1}^{3}\gamma_{i}\beta^{i}.
\label{(2.21)}
\end{equation}
In the course of this first part of the analysis one assumes that the gradient of pressure
vanishes at the center of gravity of each body, which implies that the motion remains geodesic
and hence it is legitimate to limit ourselves to the consideration of unbundled media \cite{CM6}.

\subsection{Physics of gravitating bodies with finite extension}

The Einstein modification ${\cal D}$ in (2.21) contains the functions 
$\gamma,\gamma_{i},\zeta$ obtained by integrating over all
bodies, and hence is related to finite size and internal structure of such bodies, that we will later 
identify with Sun, Earth, Moon, all planets with their satellites, and a mechanical satellite
sent off from the Earth. Following Ref. \cite{CM6}, for each body we can think of
the domain $S$ as the disjoint union of $C$ with $S'$, where $C$ denotes the domain occupied by the body,
while $S'$ is the residual portion of $S$. We shall use hereafter the notation according to which
\begin{equation}
\gamma' \equiv \left . \gamma \right |_{S'} , \;
\gamma'' \equiv \left . \gamma \right |_{C},
\label{(2.22)}
\end{equation}
\begin{equation}
\zeta' \equiv \left . \zeta \right |_{S'} , \;
\zeta'' \equiv \left . \zeta \right |_{C},
\label{(2.23)}
\end{equation}
so that we can write the decompositions
\begin{equation}
\gamma=\gamma'+\gamma'', \; \zeta=\zeta'+\zeta'',
\label{(2.24)}
\end{equation}
and also, for the two parts (Newton and Einstein) of the full Lagrangian, the resulting splits
\begin{equation}
{\cal N}={\cal N}'+{\cal N}'', \; 
{\cal D}={\cal D}'+{\cal D}'',
\label{(2.25)}
\end{equation}
where ${\cal N}',{\cal D}'$ are the part that we would have if the body $C$ were suppressed, whereas
${\cal N}'',{\cal D}''$ characterize the influence of the body $C$ on the motion of a point $P \in C$.
The explicit formulas we need are
\begin{equation}
{\cal N}'={1 \over 2}\beta^{2}+\gamma', \;
{\cal N}''=\gamma'',
\label{(2.26)}
\end{equation}
\begin{equation}
{\cal D}'={1 \over 2}{{\cal N}'}^{2}-{\gamma'}^{2}+\zeta'
+\gamma' \beta^{2}-4 \sum_{i=1}^{3}\gamma_{i}' \beta^{i},
\label{(2.27)}
\end{equation}
\begin{equation}
{\cal D}''={\cal N}'\gamma''-{1 \over 2}{\gamma ''}^{2}-2 \gamma' \gamma''
+\varphi'' + \psi'' + \nu''
+\gamma'' \beta^{2} -4 \sum_{i=1}^{3}\gamma_{i}'' \beta^{i}.
\label{(2.28)}
\end{equation}
Indeed, the internal forces resulting from $\gamma''$ are in general stronger than the external forces
resulting from the potential $\gamma'$. Nevertheless, in the equation ruling the motion of the center
of gravity, the contributions of derivatives of $\gamma''$ cancel pairwise exactly. It is here that
resides, conceptually, the origin of the cancellation principle in classical mechanics \cite{CM6}.

\subsection{Center of gravity, quasi-translational motions, size vs. distance}

At this stage, since the mere use of Eqs. (2.22)-(2.28) would lead to unmanageable equations and would not
shed enough light on the N-body problem in General Relativity, some further physical assumptions come into
play that are indeed satisfied {\it approximately} by all known celestial bodies in the solar system. 
They are as follows \cite{CM6}.
\vskip 0.3cm
\noindent
{\bf A1} The center of gravity $P_{G}$ of each body $C$ is {\it substantial}, i.e., it always adheres to the
same material element. This implies that its motion will be characterized, as it occurs for any other 
material point $P$, by a Lagrangian ${\cal L}={\cal N}+{\cal D}$, with the Einstein perturbation being
given by the term ${\cal D}$. Furthermore, we shall assume that the center of gravity $P_{G}$ is always a
center of gravitation. The latter condition means that $P_{G}$ is a point where the Newtonian attractions
of material elements of the body\footnote{Recall that the Newtonian potential $\gamma$ is bounded everywhere 
and it vanishes at infinity, hence there exists a point of minimum for $\gamma$ at which its gradient 
vanishes, which means that the force vanishes at this point.} (i.e., the internal forces) add up to zero. 
In other words, $P_{G}$ coincides with the mass center of the body.   
\vskip 0.3cm
\noindent
{\bf A2} The body performs a quasi-translational motion. Indeed, in a translational motion, all points
of the body have, at any instant $t$, the same vector speed, e.g., the speed ${\vec v}_{g}$ of the
center of gravity. We can still regard as a translation every motion for which, defining
\begin{equation}
\left | \bigtriangleup {\vec v}(t) \right | \equiv
\left | {\vec v}_{P_{i}}(t)-{\vec v}_{P_{j}}(t) \right | \; 
\forall P_{i},P_{j} \in C,
\label{(2.29)}
\end{equation}
one has always
\begin{equation}
{{\left | \bigtriangleup {\vec v} \right |}\over 
{\left | {\vec v}_{g} \right |}} <<1.
\label{(2.30)}
\end{equation}
We need sufficiently small values of the ratio in (2.30), e.g., of order $10^{-2}$, 
so that one can neglect, as a quantity of order greater than $1$, every product of the type
$$
\beta^{2}{{\left | \bigtriangleup {\vec v} \right |}\over 
{\left | {\vec v}_{g} \right |}} , \;
\gamma {{\left | \bigtriangleup {\vec v} \right |}\over 
{\left | {\vec v}_{g} \right |}} .
$$
This is precisely what happens for planetary motions. Their deformations are initially negligible
and they behave, as a consequence, as essentially rigid bodies. Their motion is actually a composition
of translation and rotation. However, for every point of the body, the speed resulting from rotation
attains only a few percent of the common speed of translation. For example, in the case of the Earth,
one has
$$
{{\left | \bigtriangleup {\vec v} \right |}\over 
{\left | {\vec v}_{g} \right |}} \approx 3 \cdot 10^{-2}.
$$
\vskip 0.3cm
\noindent
{\bf A3} On denoting by $d$ the maximal size of the body $C$, and by $R$ the minimal
Euclidean distance $d_{E}$ between points of $C$ and points of the residual portion of $S'$:
\begin{equation}
R \equiv {\rm min} \; d_{E}(P_{j}(C),P_{k}(S')),
\label{(2.31)}
\end{equation}
the quantity $\left({d \over R}\right)^{2}$ is negligible. For the Sun-Earth system, one has indeed
$\left({d \over R}\right)^{2} \approx 10^{-4}$.

By virtue of {\bf A1}, the gradient of the potential $\gamma''$ vanishes at the center of gravity, and hence
$\gamma''$ behaves as a constant. Moreover, by virtue of {\bf A2}, the dimensionless velocities $\beta^{i}$
defined in (2.9) are constant within the body $C$. On defining
\begin{equation}
\gamma'' \equiv {\widetilde \omega}={\rm constant},
\label{(2.32)}
\end{equation}
one finds from (2.10)-(2.14)
\begin{equation}
-\varphi''={\widetilde \omega}\gamma''={\widetilde \omega}^{2},
\label{(2.33)}
\end{equation}
\begin{equation}
\psi''={3 \over 2}\beta^{2}\gamma''={3 \over 2}{\widetilde \omega}\beta^{2},
\label{(2.34)}
\end{equation}
\begin{equation}
\gamma_{i}''=\beta^{i}\gamma''={\widetilde \omega}\beta^{i},
\label{(2.35)}
\end{equation}
\begin{equation}
\nu''={1 \over 2}{G_{N}\over c^{2}}{\partial^{2}\over \partial (x^{0})^{2}}
\int_{C}\mu r {\rm d}C=0, \;
r=d_{E}(P_{j},P_{G}) \; \forall P_{j} \in C.
\label{(2.36)}
\end{equation}
We stress that $\nu''$ vanishes because the integral in (2.36) is constant during the motion.

\subsection{The $\chi_{k}$ coefficients}

Hereafter we denote with $P_{k}$ the center of gravity of the body $C_{k}$,
for all $k=0,1,...,n-1$. We also denote with $l_{k}$ the gravitational radius of the 
$k$-th body having mass $m_{k}$ (we assume, on experimental ground, the equality of
inertial and gravitational mass, and also of active and passive gravitational mass), i.e.,
\begin{equation}
l_{k} \equiv {G_{N}m_{k}\over c^{2}},
\label{(2.37)}
\end{equation}
which, in the solar system, does not exceed the $1.5$ km for the gravitational radius of the Sun.
The assumption {\bf A3} implies that the potential of the body $C_{k}$ acting on the center of gravity $P_{h}$
of $C_{h}$ is given, as if the mass of $C_{k}$ were completely concentrated at the center
of gravity $P_{k}$, in the form ${G_{N}m_{k}\over r(P_{k},P_{h})}$, where the denominator is the
Euclidean distance $d_{E}(P_{k},P_{h})$. The dimensionless form of such a potential is obtained
dividing by $c^{2}$, i.e.
$$
{G_{N}m_{k} \over c^{2} r(P_{k},P_{h})}={l_{k}\over r(P_{k},P_{h})}.
$$
Now we need extra labels in the notation, since we are going to derive the Lagrangian 
in the form ${\cal N}+{\cal D}$ for each 
(celestial) body. For this purpose, following again Ref. \cite{CM6}, 
we shall denote by $\beta_{h}^{2}$ the square of the velocity of $P_{h}$, by $(\beta_{h})_{i}$
the component along the axis $x^{i}$ of $\beta_{h}$, and by $\gamma_{h}'$ the potential at
$P_{h}$ resulting from all bodies $C_{k}$ described by an index $k \not =h$. This latter condition
means that
\begin{equation}
\gamma_{h}'=\sum_{k=0}^{n-1}(1-\delta_{k,h}){l_{k}\over r(P_{k},P_{h})},
\label{(2.38)}
\end{equation}
with the understanding that the Kronecker $\delta$ plays the role of giving vanishing weight
to the divergent term ${l_{k}\over r(P_{k},P_{k})}$, which is therefore ruled out from the sum
(sometimes this is expressed by the $\sum'$ notation, which is here made clearer).

In light of (2.32), we can regard as being constant the integral
\begin{equation}
\gamma_{h}''(P)={G_{N}\over c^{2}}\int_{C_{h}}
{\mu(Q)\over r(Q,P)}{\rm d}C_{h}={\widetilde \omega}_{h},
\label{(2.39)}
\end{equation}
which is the potential of the body $C_{h}$ at any point $P$ of $C_{h}$ itself. The
constant ${\widetilde \omega}_{h}$ is majorized by the ratio ${{\overline l}\over {\overline r}}$,
where ($\partial C_{h}$ being the boundary of $C_{h}$)
\begin{equation}
{\overline r} \equiv {\rm max} \; d_{E}(P_{h},\partial C_{h}),
\label{(2.40)}
\end{equation}
while ${\overline l}\equiv {G_{N}{\overline m}\over c^{2}}$, ${\overline m}$ being 
the mass contained within a homogeneous sphere of density 
${\overline \mu}$ and radius ${\overline r}$, having set
\begin{equation}
{\overline \mu}={\rm sup}(\mu) \; {\rm within} \; C_{h}.
\label{(2.41)}
\end{equation}
Equations (2.39)-(2.41) tell us that ${\widetilde \omega}$ is a quantity of
first order, being close to a quantity proportional to a Newtonian potential.
This simple property will be nicely exploited below.

Next, we consider an integration domain $S'$ consisting of all bodies $C_{k}$ with the
exception of the body $C_{h}$, in formulas
\begin{equation}
S' \equiv \cup_{k}C_{k}-C_{h},
\label{(2.42)}
\end{equation}
and, for points $Q \in S'$ and $P \in C_{h}$, we consider the decomposition of the function
$\varphi_{h}$ for the body $C_{h}$ in the form
\begin{equation}
\varphi_{h}=\varphi_{h}'+\varphi_{h}'',
\label{(2.43)}
\end{equation}
where 
\begin{equation}
\varphi_{h}'=-{G_{N}\over c^{2}} \int_{S'}{\mu \gamma \over r(Q,P)}{\rm d}S'.
\label{(2.44)}
\end{equation}
In light of assumption {\bf A3}, we can re-express (2.44) in the form
\begin{eqnarray}
\varphi_{h}' &=& -{G_{N}\over c^{2}}
\sum_{k=0}^{n-1}{(1-\delta_{k,h})\over r(P_{k},P)} \int_{C_{k}}
\mu(Q)\gamma_{k}(Q){\rm d}C_{k} 
\nonumber \\
&=& -{G_{N}\over c^{2}} \sum_{k=0}^{n-1}{(1-\delta_{k,h})\over r(P_{k},P)} 
\left( \sum_{j=0}^{n-1} (1-\delta_{j,k})\dfrac{l_j}{r(P_j,P_k)} \int_{C_k} \mu(Q) {\rm d}C_k 
+ \dfrac{G_N}{c^2} \int_{C_k} \mu(Q) {\rm d}C_k \int_{C_k} \dfrac{\mu(Q^\prime)}{r(Q,Q^\prime)} {\rm d}C_k  \right) 
\nonumber \\
&=& - \sum_{k=0}^{n-1}(1-\delta_{k,h})\dfrac{l_k}{ r(P_{k},P)} 
\sum_{j=0}^{n-1}(1-\delta_{j,k})\dfrac{l_j}{ r(P_{j},P_{k})} 
- \sum_{k=0}^{n-1}(1-\delta_{k,h})\dfrac{l_k \chi_k}{ r(P_{k},P)},
\label{(2.45)}
\end{eqnarray}
having defined \cite{CM6}
\begin{equation}
\chi_{k} \equiv {1 \over l_{k}}\left({G_{N}\over c^{2}}\right)^{2}
\int_{C_{k}}\mu(Q) {\rm d}C_{k} \int_{C_{k}} {\mu(Q')\over r(Q',Q)}{\rm d}C_{k}
= \dfrac{G_N}{c^2} \widetilde{\omega}_k= {\rm constant},
\label{(2.46)}
\end{equation}
and where each $\gamma_{k}$ potential in (2.45) has been split as
in (2.24), i.e., $\gamma_{k}'(Q)+\gamma_{k}''(Q)$, and use of (2.38) and (2.39) has been made in order
to express $\gamma_{k}'(Q)$ and $\gamma_{k}''(Q)$, respectively. The existence of the constant 
coefficients $\chi_{k}$ is conceptually interesting, but their 
values are extremely small, because the integrals occurring in (\ref{(2.46)}) are finite but are 
multiplied by the square of the ratio ${G_{N}\over c^{2}}$.

\subsection{The effacement property}

At this stage, one can obtain the desired decomposition of the Lagrangian for the $h$-th body $C_{h}$ in the
form \cite{CM6}
\begin{equation}
{\cal L}_{h}={\cal N}_{h}+{\cal D}_{h}'+{\cal D}_{h}'',
\label{(2.47)}
\end{equation}
where ${\cal N}_{h}$ is the Newtonian term
\begin{equation}
{\cal N}_{h}={1 \over 2}(\beta_{h})^{2}+\gamma_{h}'={\cal N}_{h}',
\label{(2.48)}
\end{equation}
while ${\cal D}_{h}'$ is the pointlike Einstein perturbation
\begin{equation}
{\cal D}_{h}'={1 \over 2}({\cal N}_{h}')^{2}- (\gamma_{h}')^{2}
+\zeta_{h}'+\gamma_{h}'(\beta_{h})^{2}
-4 \sum_{i=1}^{3}(\Gamma_{h}')_{i}(\beta_{h})_{i},
\label{(2.49)}
\end{equation}
$(\beta_{h})_{i}$ denoting the $i$-th component of the velocity of the $h$-th body
as we said before (2.38), and
\begin{equation}
(\Gamma_{h}')_{i} \equiv \sum_{k=0}^{n-1}(1-\delta_{k,h})
{l_{k}(\beta_{k})_{i}\over r(P_{k},P_{h})}.
\label{(2.50)}
\end{equation}
The expression (\ref{(2.47)}) of the Lagrangian is completed by ${\cal D}_{h}''$, i.e. the 
Einstein perturbation resulting from the extension of bodies.
Upon defining
\begin{equation}
m \equiv \sum_{k=0}^{n-1} m_{k}, \; l \equiv {G_{N}m \over c^{2}}, \;
l_{k} \equiv {G_{N}m_{k}\over c^{2}}, \;
\lambda_{k} \equiv {m_{k}\over m}={l_{k}\over l},
\label{(2.51)}
\end{equation}
the perturbation ${\cal D}_{h}''$ reads as
\begin{equation}
{\cal D}_{h}''=-{\widetilde \omega}_{h}(\beta_{h})^{2}-l \sum_{k=0}^{n-1}(1-\delta_{k,h})
{\lambda_{k}(\chi_{k}+2 {\widetilde \omega}_{h})\over r(P_{k},P_{h})}.
\label{(2.52)}
\end{equation}

We can now recall that Lagrangians differing by a multiplicative constant give rise to equivalent
equations of motion. The simple and profound idea of Levi Civita was to consider a
first-order quantity $\sigma_{h}$, and to multiply ${\cal L}_{h}$ by $(1+\sigma_{h})$.
After doing this, one can try to choose $\sigma_{h}$ in such a way that the occurrence
of the constant ${\widetilde \omega}_{h}$ gets exactly cancelled. This is indeed feasible
because, up to higher order terms here negligible, one finds
\begin{equation}
(1+\sigma_{h}){\cal L}_{h} \sim (1+\sigma_{h}){\cal N}_{h}
+{\cal D}_{h}'+{\cal D}_{h}'',
\label{(2.53)}
\end{equation}
where, in particular,
\begin{equation}
(1+\sigma_{h}){\cal N}_{h}+{\cal D}_{h}''={1 \over 2}(\beta_{h})^{2}
+\left({1 \over 2}\sigma_{h}-{\widetilde \omega}_{h} \right)(\beta_{h})^{2}
+l \sum_{k=0}^{n-1}(1-\delta_{k,h})
{\lambda_{k}[(1-\chi_{k})+(\sigma_{h}-2{\widetilde \omega}_{h})] \over r(P_{k},P_{h})}.
\label{(2.54)}
\end{equation}
This formula suggests choosing
\begin{equation}
\sigma_{h}=2 {\widetilde \omega}_{h},
\label{(2.55)}
\end{equation}
to achieve the desired cancellation. The result is also consistent with what we know already
about the first-order nature of the constant ${\widetilde \omega}_{h}$. We can further define
\begin{equation}
\Lambda_{k} \equiv \lambda_{k}(1-\chi_{k}),
\label{(2.56)}
\end{equation}
Note also that the pointlike Einstein perturbation ${\cal D}_{h}'$ is still expressed in terms of
the dimensionless $\lambda_{k}$ coefficients, but we can insert also therein the $\Lambda_{k}$
defined in (2.56), because
\begin{equation}
{\cal D}_{h}'(\lambda_{k})={\cal D}_{h}'(\Lambda_{k})+{\rm higher}-{\rm order}
\; {\rm terms}.
\label{(2.57)}
\end{equation}
One finds therefore that each (celestial) body is ruled by a pointlike Lagrangian $L_{h}$
where the Einstein perturbation is no longer split into pointlike plus finite-size part, and
one can write \cite{CM6}
\begin{equation}
L_{h}={\cal N}_{h}+{\cal D}_{h},
\label{(2.58)}
\end{equation}
where, having defined
\begin{equation}
\gamma_{h} \equiv l \sum_{k=0}^{n-1}(1-\delta_{k,h}){\Lambda_{k}\over r(P_{k},P_{h})},
\label{(2.59)}
\end{equation}
\begin{equation}
(\Gamma_{h})_{i} \equiv l \sum_{k=0}^{n-1}(1-\delta_{k,h})
{\Lambda_{k}(\beta_{k})_{i}\over r(P_{k},P_{h})},
\label{(2.60)}
\end{equation}
\begin{equation}
\zeta_{h} \equiv \varphi_{h}+\psi_{h}+\nu_{h},
\label{(2.61)}
\end{equation}
\begin{equation}
\varphi_{h} \equiv -l^{2}\sum_{k=0}^{n-1}\left[(1-\delta_{k,h})
{\Lambda_{k}\over r(P_{k},P_{h})} 
\sum_{s=0}^{n-1} (1-\delta_{s,k})
{\Lambda_{s} \over r(P_{s},P_{k})}\right],
\label{(2.62)}
\end{equation}
\begin{equation}
\psi_{h} \equiv {3 \over 2} l \sum_{k=0}^{n-1} (1-\delta_{k,h})
{\Lambda_{k}(\beta_{k})^{2} \over r(P_{k},P_{h})},
\label{(2.63)}
\end{equation}
\begin{equation}
\nu_{h} \equiv {1 \over 2}l {\partial^{2}\over \partial (x^{0})^{2}}
\sum_{k=0}^{n-1}(1-\delta_{k,h})  
\Lambda_{k} r(P_{k},P_{h}),
\label{(2.64)}
\end{equation}
the Newtonian term takes the familiar form
\begin{equation}
{\cal N}_{h} \equiv {1 \over 2}(\beta_{h})^{2}+\gamma_{h},
\label{(2.65)}
\end{equation}
while the Einstein perturbation is eventually expressed by the sum of functions
\begin{equation}
{\cal D}_{h} \equiv {1 \over 2}({\cal N}_{h})^{2}-(\gamma_{h})^{2}+\zeta_{h}
+\gamma_{h}(\beta_{h})^{2}-4 \sum_{i=1}^{3}(\Gamma_{h})_{i}(\beta_{h})_{i}.
\label{(2.66)}
\end{equation}
Equations (2.58)-(2.66) lead to an accurate scheme for writing down and studying the
equations of motion of each (celestial) body, and provide a precise statement of the
cancellation principle in General Relativity: on going from Newtonian to relativistic
celestial mechanics, the effects of extension and internal structure of bodies are encoded
in the family of $\Lambda_{k}$ parameters, Eq. (2.56), which differ only by a tiny amount
(see Eq. (2.46)) from the dimensionless mass ratios ${m_{k}\over m}$
of Eq. (2.51). Thus, the effects
of finite extension of bodies get eventually dissolved neatly, and it is as if we were dealing
with material points which do not affect at all their center of gravity. This holds also
for the solar system. 

As is stressed in Ref. \cite{CM14}, a more rigorous proof of such a cancellation principle
can be found in Ref. \cite{CM5}, to which we refer the reader interested in a broader
understanding.

\section{Sun-Earth, Sun-Earth-Moon and $N$-body dynamics}

Driven by the concepts outlined in the previous sections, we now aim at investigating the system 
consisting of the Sun, Earth and as many additional celestial bodies as possible 
 by means of the Lagrangian (\ref{(2.58)}). 
An important comment should be made at this stage. In fact, in the most general case,   
acceleration terms appear in (\ref{(2.58)}). However, 
by bearing in mind that $\chi_k \ll 1$ and $\lambda_k = {\rm constant}$ 
(cf. Eqs. (\ref{(2.46)}) and (\ref{(2.51)})), it easily follows that 
\begin{equation}
\nu_h = 0, \; \; \; \; \; \; \forall h, 
\label{(3.1)}
\end{equation}
since the Euclidean distance $r (P_k,P_h)$ occurring in Eq. (\ref{(2.64)}) depends on time only implicitly, 
through the coordinates $x^i(t)$, i.e., 
\begin{equation}
\dfrac{\partial}{\partial t} r (P_k,P_h)=\dfrac{\partial}{\partial t} r \left(x^1(t),x^2(t),x^3(t)\right)=0.
\label{(3.2)}
\end{equation}
In other words, under our assumption the Lagrangian (\ref{(2.58)}) turns out to be a function of the 
Euclidean coordinates $x^1(t),x^2(t),x^3(t)$ and their first time derivatives only. Moreover, Eq. 
(\ref{(3.1)}) is valid also in the more general case of a Lagrangian function depending explicitly on 
the time variable $t$, since the value of the integral 
\begin{equation}
\int_{{\rm D}} \mu r \, {\rm d} C,
\label{(3.3)}
\end{equation}
evaluated in the region ${{\rm D}}$ made up of all those spatial points which are very 
distant from $P_h$ reduces nearly to zero because it turns out to be of order $\left({d \over R}\right)^{2}$, 
which, according to hypothesis {\bf A3}, represents a negligible quantity.

\begin{figure} [htbp] 
\includegraphics[scale=0.7]{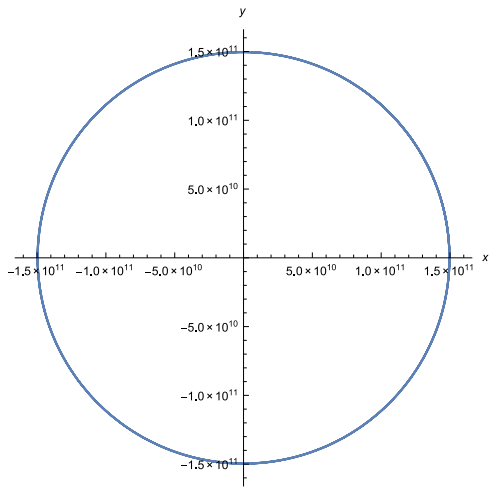}
\caption{Circular orbit around the Sun of a body having the mass of the Earth as 
described by the Lagrangian function (\ref{(2.58)}). 
The same orbit is obtained also when Eqs. (\ref{(3.4)})--(\ref{(3.6)}) are employed.}
\label{Fig1.eps}
\end{figure}

\begin{figure} [htbp] 
\includegraphics[scale=0.9]{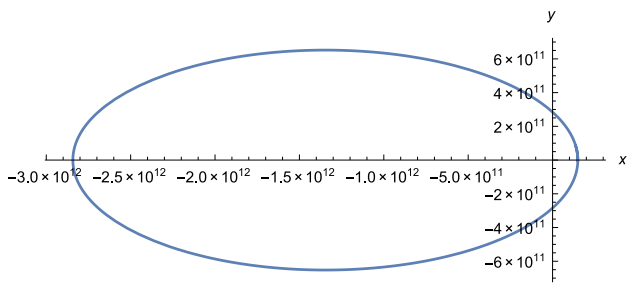} 
\caption{Elliptic orbit around the Sun of a body having the mass of the Earth as described 
by the Lagrangian function (\ref{(2.58)}). 
The same orbit is obtained also when Eqs. (\ref{(3.4)})--(\ref{(3.6)}) are employed.}
\label{Fig2.eps}
\end{figure}

As a first step, we have analysed the system made up of just two bodies by considering the case 
of one celestial body orbiting a fixed massive object (i.e., the Sun). We have found that 
perihelion shift predicted by the Levi-Civita Lagrangian is in accordance with the well-known results expected 
within the usual 1PN picture of general relativity. In fact by employing the latter approximation, 
the orbit of the revolving body (in the equatorial plane $\theta= \pi/2$) 
is described by the well-known relations \cite{MTW}
\begin{equation}
r(\phi)= \dfrac{(1-e^2)a}{1+e \cos \left[ \left(1-\delta \phi_0/2 \pi \right) \phi \right]},
\label{(3.4)}
\end{equation}
and
\begin{equation}
\begin{split}
& x= e\, a + r(\phi) \cos \phi,\\
& y= r(\phi) \sin \phi,
\end{split}
\label{(3.5)}
\end{equation}
with
\begin{equation}
\delta \phi_0 = \dfrac{6 \pi G_N M}{c^2 a(1-e^2)},
\label{(3.6)}
\end{equation}
$e$ being the eccentricity of the orbit of the orbiting body, $a$ the semi-major axis and $M$ 
the mass of the massive object. As demonstrated in Figs. \ref{Fig1.eps} and \ref{Fig2.eps}, 
the outcomes achieved within the Levi-Civita framework are in agreement with the ones expected 
through the 1PN approximation method, witnessing that Levi-Civita actually made a mistake in 
Ref. \cite{CM6} when concluding that his pattern predicts a more pronounced shift of the perihelion 
in the orbit of the revolving body. This point is in accordance with the analysis of Ref. \cite{CM14}. 

Moreover, the Newtonian relations expressing the eccentricity $e$ of the orbit as a function of the initial 
velocity $v_0$ of the revolving body and its distance $\mathcal{R}$ from the massive one according to
\begin{equation}
\begin{split}
& v_0 = \sqrt{\dfrac{G_N M}{\mathcal{R}}}, \;\;\;\; \; \;    \;\;\;\; \; \;\;\;\;\; \; \;\;\;\;\; \; \; \;\;\;\; 
\; \;\;\;\;\; \; \;\;\;\;\; \; \;\;\;\; {\rm circular \; orbit}, \\
& 0 < v_0 < \sqrt{\dfrac{2 G_N M}{\mathcal{R}}},\;\;\;  \left(v_0  \neq \sqrt{\dfrac{G_N M}{\mathcal{R}}}\right), 
\;\;\;\; \; \; {\rm elliptic \; orbit}, \\
& v_0 = \sqrt{\dfrac{2 G_N M}{\mathcal{R}}}, \;\;\;\; \; \;     \;\;\;\; \; \;\;\;\;\; \; \;\;\;\;\; \; \; \;\;\;\; 
\; \;\;\;\;\; \; \;\;\;\;\; \; \;\;\;{\rm parabolic \; orbit},\\
& v_0 > \sqrt{\dfrac{2 G_N M}{\mathcal{R}}}, \;\;\;\; \; \;      \;\;\;\; \; \;\;\;\;\; \; \;\;\;\;\; \; \; \;\;\;\; 
\; \;\;\;\;\; \; \;\;\;\;\; \; \;\;\; {\rm hyperbolic \; orbit},
\end{split}
\label{(3.7)}
\end{equation}
are found to be respected by applying the Levi-Civita framework. 

\begin{figure}
\includegraphics[scale=0.7]{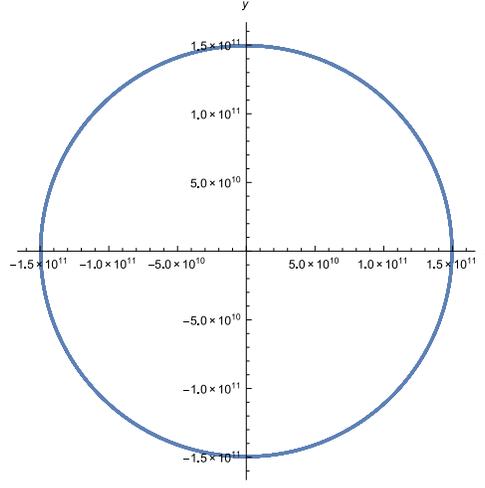}
\caption{Circular orbit around the Sun in the gravitational field produced by the Sun itself and the Earth 
of a body having the mass of the Moon as described by 
the Levi-Civita Lagrangian (\ref{(2.58)}). Note that this Figure is similar to Fig. \ref{Fig1.eps} since the 
distances of the Earth and the Moon from the Sun are quite the same.}
\label{Fig3.eps}
\end{figure}

After that, the system consisting of three bodies (i.e, the Earth and the Moon orbiting the Sun) has been considered. 
We have recovered the orbit of the Moon around the Earth and around the Sun. Once again, the Levi-Civita Lagrangian produces 
negligible differences with respect to both the 1PN model and Newtonian theory. As an example, Fig. \ref{Fig3.eps} 
describes the motion around the Sun (and in the presence of the Earth describing the usual Newtonian 
elliptic orbit around the Sun) of a body having the same mass as the Moon 
in the hypothesis of circular motion. Also the cases of elliptic, parabolic, and hyperbolic orbits have been analysed.

The presence in our model of such periodic solutions will become crucial at the end of Sec. IV
(see also appendix B and the remarks therein on the Sturm-Liouville problem). 

Many complications arise when we deal with the Sun-Jupiter-Earth-Moon-planetoid system so that some 
further simplification must be assumed for the Lagrangian function (\ref{(2.58)}). If we suppose that the 
velocity $\beta_k$ of the Sun, Earth, Moon, and Jupiter is negligible if compared to the planetoid velocity, 
we can then set (cf. Eqs. (\ref{(2.60)}) and (\ref{(2.63)}))
\begin{equation}
\begin{split}
& \left( \Gamma_h \right)_{i} =0, \\
& \psi_h =0, 
\end{split}
\label{(3.8)}
\end{equation}  
where the index $h$ must be understood as labelling the planetoid. In this way the resulting Lagrangian is such that
\begin{equation}
{\cal D}_{h} = {1 \over 2}({\cal N}_{h})^{2}-(\gamma_{h})^{2}+\varphi_{h}
+\gamma_{h}(\beta_{h})^{2},
\label{(3.9)}
\end{equation}
and hence it assumes a form that can be studied more easily. 

\subsection{Sun-Earth system}

In the simple system made up of just the Earth orbiting the Sun, the joint effect of Eqs. (\ref{(2.58)}) 
and (\ref{(3.9)}) gives rise to a Lagrangian function describing the planar motion of the Earth
\begin{equation}
m_E c^2\mathcal{L}_{E}(x,y,\dot{x},\dot{y})= m_E c^2\left(\mathcal{N}_E + \mathcal{D}_E\right),
\label{(3.10)}
\end{equation}
($m_E$ being the mass of the Earth) such that 
\begin{equation}
\mathcal{N}_E= \dfrac{1}{2} \left(\beta_E\right)^2 + \gamma_E=\dfrac{1}{2c^2} \left( \dot{x}^2+\dot{y}^2 \right) 
+ \dfrac{G m_S}{c^2} \dfrac{1}{r},
\label{(3.11)}
\end{equation}
\begin{equation}
\mathcal{D}_E=\dfrac{1}{2} \left(\mathcal{N}_E \right)^2- \left(\gamma_E \right)^2 
+ \varphi_E + \gamma_E \left( \beta_E \right)^2,
\label{(3.12)}
\end{equation}
with (cf. (\ref{(2.62)}))
\begin{equation}
\varphi_E= - \dfrac{l^2}{r^2} \Lambda_S \Lambda_E,
\label{(3.13)}
\end{equation}
where $m_S$ is the mass of the Sun and 
\begin{equation}
r=r(t)=\sqrt{x^{2}(t)+y^{2}(t)},
\label{(3.14)}
\end{equation}
denoting the Euclidean planar distance between the point $\bold{r}(t)=(x(t),y(t))$ occupied at the time 
$t$ by the Earth and the Sun, supposed at rest at the origin of the coordinate system. The resulting 
Euler-Lagrange equations (which, once integrated numerically, have led to Figs. 
\ref{Fig1.eps} and \ref{Fig2.eps}) can be written in the form
\begin{equation}
\renewcommand{\arraystretch}{2.0}
\begin{dcases}
& f(x,y,\dot{x},\dot{y}) \ddot{x}+ g(\dot{x},\dot{y}) \ddot{y}+ h_1(x,y,\dot{x},\dot{y}) =0, \\
& g(\dot{x},\dot{y}) \ddot{x} + f(x,y,\dfrac{\dot{x}}{\sqrt{3}},\sqrt{3}\dot{y}) \ddot{y} +  h_2(x,y,\dot{x},\dot{y}) =0, 
\end{dcases} \\ [2em]
\label{(3.15)}
\end{equation}
where
\begin{equation}
f\left(x,y,\dot{x},\dot{y}\right)= \dfrac{1}{2}m_E \left(2+\dfrac{6 l \Lambda_S}{r} + \dfrac{3 \dot{x}^2+\dot{y}^2}{c^2} \right),
\label{(3.16)}
\end{equation}
\begin{equation}
g\left(\dot{x},\dot{y}\right)=\frac{m_E }{c^2} \dot{x} \, \dot{y},
\label{(3.17)}
\end{equation}
\begin{equation}
h_1(x,y,\dot{x},\dot{y})= \dfrac{l m_E \Lambda_S \left[ -4c^2 l \Lambda_E r x - 2c^2 l \Lambda_S r x 
+ r^2 \left(-6 y \dot{x} \dot{y} + 2 c^2 x -3 x\dot{x}^2 + 3 x \dot{y}^2 \right)\right] }{2 r^5},
\label{(3.18)}
\end{equation}
\begin{equation}
h_2(x,y,\dot{x},\dot{y})= \dfrac{l m_E \Lambda_S \left[ -4c^2 l \Lambda_E r y - 2c^2 l \Lambda_S r y 
+ r^2 \left(-6 x \dot{x} \dot{y} + 2 c^2 y +3 x\dot{x}^2 - 3 y \dot{y}^2 \right)\right] }{2 r^5}.
\label{(3.19)}
\end{equation}
Note that it is possible to obtain $h_2(x,y,\dot{x},\dot{y})$ starting from $h_1(x,y,\dot{x},\dot{y})$ 
(or vice-versa) by setting
\begin{equation}
\begin{split}
& x\longleftrightarrow y, \\
& \dot{x}^2 \rightarrow - \dot{x}^2, \\
& \dot{y}^2 \rightarrow - \dot{y}^2,
\end{split}
\label{(3.20)}
\end{equation}
whereas the terms linear in the velocities have the freedom to change according to
\begin{equation}
\begin{split}
& \dot{x} \rightarrow  \pm \dot{x}, \\
& \dot{y} \rightarrow \pm \dot{y}.
\end{split}
\label{(3.21)}
\end{equation}

\subsection{Sun-Earth-Moon system}

If we analyse the Sun-Earth-Moon system, the Lagrangian for the Moon becomes
\begin{equation}
m_M c^2\mathcal{L}_{M}(x,y,\dot{x},\dot{y})= m_M c^2\left(\mathcal{N}_M + \mathcal{D}_M\right),
\label{(3.22)}
\end{equation}
$m_M$ being the mass of the Moon, while
\begin{equation}
\gamma_M = \dfrac{l_S}{r}+\dfrac{l_E}{r_{EM}},
\label{(3.23)}
\end{equation}
\begin{equation}
\varphi_M = -l^2 \left[\dfrac{\Lambda_S}{r} \left( \dfrac{\Lambda_E}{r_E}
+\dfrac{\Lambda_M}{r} \right)+\dfrac{\Lambda_E}{r_{EM}} \left( \dfrac{\Lambda_M}{r_{EM}}
+\dfrac{\Lambda_S}{r_E} \right) \right],
\label{(3.24)}
\end{equation}
where now $r=r(t)=\sqrt{x^{2}(t)+y^{2}(t)}$ denotes the distance of the Moon from the Sun, 
$r_{EM}=\vert \bold{r}-\bold{r}_E\vert =\sqrt{\left(x_E(t)-x(t)\right)^2+\left(y_E(t)-y(t)\right)^2}$ 
represents the distance between the Earth, having coordinates $\bold{r}_E(t)=(x_E(t),y_E(t))$, 
and the Moon, whereas $r_E$ is the distance between the Sun and the Earth. The resulting Euler-Lagrange 
equations, giving rise to Fig. \ref{Fig3.eps}, assume the form
\begin{equation}
\renewcommand{\arraystretch}{2.0}
\begin{dcases}
& \tilde{f}(x,y,x_E,y_E,\dot{x},\dot{y}) \ddot{x}+ \tilde{g}(\dot{x},\dot{y}) \ddot{y}
+ \tilde{h}_1(x,y,x_E,y_E,\dot{x},\dot{y}) =0, \\
& \tilde{g}(\dot{x},\dot{y}) \ddot{x} + \tilde{f}(x,y,x_E,y_E,\dfrac{\dot{x}}{\sqrt{3}},
\sqrt{3}\dot{y}) \ddot{y} +  \tilde{h}_2(x,y,x_E,y_E,\dot{x},\dot{y})=0, 
\end{dcases} \\ [2em]
\label{(3.25)}
\end{equation}
with
\begin{equation}
\tilde{f}(x,y,x_E,y_E,\dot{x},\dot{y})= \dfrac{m_M}{2c^2} \left( 2c^2 + \dfrac{6 c^2 l_S}{r} 
+ \dfrac{6 c^2 l_E}{r_{EM}}+ 3 \dot{x}^2 + \dot{y}^{2} \right),
\label{(3.26)}
\end{equation}
\begin{equation}
\tilde{g}(\dot{x},\dot{y})= \dfrac{m_M \dot{x} \dot{y}}{c^2},
\label{(3.27)}
\end{equation}
\begin{equation}
\begin{split}
\tilde{h}_1(x,y,x_E,y_E,\dot{x},\dot{y}) & = \dfrac{1}{2}m_M \Biggl \{\dfrac{1}{r_{EM}^3} 
\biggl (\dfrac{2c^2l_E^2x_E}{r_{EM}}+ \dfrac{4 c^2 l^2 \Lambda_E \Lambda_M x_E}{r_{EM}}
+2 c^2 l_E x_E + 2c^2l^2 \Lambda_E \Lambda_S (x_E/r_E) \\
& + \dfrac{2c^2 l_E l_S x_E}{r} + 3 l_E x_E \dot{x}^2 + 6 l_E y_E \dot{x} \dot{y} - 3 l_E x_E \dot{y}^2 \biggr) 
+ 6 y \dot{x} \biggl[ \dfrac{l_E \omega_E x_E}{r_{EM}^3} 
- \dot{y} \Bigl(\dfrac{l_S}{r^3}+\dfrac{l_E}{r_{EM}^3} \Bigr) \biggr] \\
& + x \biggl[2 c^2 \biggl ( -\dfrac{l_S^2}{r^4} - \dfrac{2 l^2 \Lambda_M \Lambda_S}{r^4} 
- \dfrac{l^2 \Lambda_E \Lambda_S }{r_E \, r^3}-\dfrac{l_E^2}{r_{EM}^4} - \dfrac{2 l^2 \Lambda_E 
\Lambda_M}{r_{EM}^4}+ \dfrac{l_E}{r_{EM}^3} - \dfrac{l^2 \Lambda_E \Lambda_S}{r_E \, r_{EM}^3} \\ 
& + \dfrac{l_S}{r^3} + 
\dfrac{l_S l_E}{r \,r_{EM}^3} -\dfrac{l_S l_E}{r^3 r_{EM}} \biggr) -\dfrac{6 l_E \omega_E y \dot{x}}
{r_{EM}^3} + 3(\dot{y}^2-\dot{x}^2) \biggl( \dfrac{ l_S}{r^3} +\dfrac{l_E}{r_{EM}^3} \biggr) \biggr] \Biggr\},
\end{split}
\label{(3.28)}
\end{equation}
$\omega_E$ being the classical pulsation associated to the motion of the Earth around the Sun 
and, likewise the two-body case, 
\begin{equation}
\tilde{h}_1(x,y,x_E,y_E,\dot{x},\dot{y})\longleftrightarrow \tilde{h}_2 (x,y,x_E,y_E,\dot{x},\dot{y}),
\label{(3.29)}
\end{equation}
if
\begin{equation}
\begin{split}
& x\longleftrightarrow y, \\
& x_E \longleftrightarrow y_E , \\
& \dot{x} \rightarrow  - \dot{x}, \\
& \dot{y} \rightarrow -\dot{y}, \\  
& \dot{x}^2 \rightarrow - \dot{x}^2, \\
& \dot{y}^2 \rightarrow - \dot{y}^2.
\end{split}
\label{(3.30)}
\end{equation}

\subsection{Case of the $h$-th body immersed in the gravitational field produced by the other $(n-1)$ bodies}

Driven by the features of the above analysis, it is possible to infer the presence of a recursive scheme 
according to which the Lagrange equations resulting from Eqs. (\ref{(2.58)}) and (\ref{(3.9)}), and describing 
the motion of the $h$-th body immersed in the gravitational field produced by the other $n-1$ bodies, are given by
\begin{equation}
\renewcommand{\arraystretch}{2.0}
\begin{dcases}
& \mathcal{F}(x_h,y_h,x_k,y_k,\dot{x}_h,\dot{y}_h) \ddot{x}_h+ \mathcal{G}(\dot{x}_h,\dot{y}_h) 
\ddot{y}_h+ \mathcal{H}_1(x_h,y_h,x_k,y_k,\dot{x}_h,\dot{y}_h)  =0, \\
& \mathcal{G}(\dot{x}_h,\dot{y}_h) \ddot{x}_h + \mathcal{F}(x_h,y_h,x_k,y_k,\dfrac{\dot{x}_h}{\sqrt{3}},
\sqrt{3}\dot{y}_h) \ddot{y}_h +  \mathcal{H}_2 (x_h,y_h,x_k,y_k,\dot{x}_h,\dot{y}_h) =0, 
\end{dcases} \\ [2em]
\; \; \; \; (k=0,1,\dots,n-1), (k \neq h),
\label{(3.31)}
\end{equation}
possessing the property
\begin{equation}
\mathcal{H}_1(x_h,y_h,x_k,y_k,\dot{x}_h,\dot{y}_h) \longleftrightarrow \mathcal{H}_2(x_h,y_h,x_k,y_k,\dot{x}_h,\dot{y}_h),  
\label{(3.32)}
\end{equation}
when
\begin{equation}
\begin{split}
& x_h\longleftrightarrow y_h, \\
& x_k \longleftrightarrow y_k , \\
& \dot{x}_h \rightarrow  - \dot{x}_h, \\
& \dot{y}_h \rightarrow -\dot{y}_h, \\  
& \dot{x}_h^2 \rightarrow - \dot{x}_h^2, \\
& \dot{y}_h^2 \rightarrow - \dot{y}_h^2.
\end{split}
\label{(3.33)}
\end{equation}

\section{Linear system of ordinary differential equations associated with the nonlinear equations of motion}

The expressions describing the relativistic motion of a massive object we have derived in the previous 
section (cf. Eqs. (\ref{(3.15)}), (\ref{(3.25)}), and (\ref{(3.31)})) clearly represent a coupled second-order 
system of nonlinear ordinary differential equations. All the coefficients occurring in these equations can 
be seen as smooth and differentiable functions on $\mathbb{R}^2$ since the distance scales occurring in our 
framework prevent the bodies from colliding. This means that our model cannot be employed to investigate the 
binary systems analysed in relativistic astrophysics 
\cite{Damour1,Damour2,Damour3,Damour4,Damour5,Damour6,Damour7,Damour8,Damour9,Damour10}, 
or those systems made up of comets or asteroids hitting a planet or other celestial bodies.

We now aim at showing that it is possible to map such systems into a second-order system 
of {\it linear} ordinary differential equations by applying the time derivative operator 
$\dfrac{{\rm d}}{{\rm d}t}$ four times to the original equations. The resulting expressions contain the 
fourth-order time derivative of the functions $x(t)$ and $y(t)$ as their unknowns.
Consider the system (\ref{(3.31)}) written in matrix form
\begin{equation}
\mathcal{A}
 \left[
 \renewcommand{\arraystretch}{2.0}
\begin{matrix}
 & \ddot{x}_h \cr
& \ddot{y}_h  
\end{matrix} 
\right]
 + 
  \left[
  \renewcommand{\arraystretch}{2.0}
 \begin{matrix}
& \mathcal{H}_1(x_h,y_h,x_k,y_k,\dot{x}_h,\dot{y}_h) \cr
 & \mathcal{H}_2(x_h,y_h,x_k,y_k,\dot{x}_h,\dot{y}_h) 
 \end{matrix}
 \right]
 =0,
\label{(4.1)}
\end{equation}
where
\begin{equation}
\mathcal{A}= 
 \left[
\renewcommand{\arraystretch}{2.0}
\begin{matrix}
& \mathcal{F}(x_h,y_h,x_k,y_k,\dot{x}_h,\dot{y}_h) &  \mathcal{G}(\dot{x}_h,\dot{y}_h) \cr
& \mathcal{G}(\dot{x}_h,\dot{y}_h) &  \mathcal{F}(x_h,y_h,x_k,y_k,\dfrac{\dot{x}_h}{\sqrt{3}},\sqrt{3}\dot{y}_h)
 \end{matrix} \right].
\label{(4.2)}
\end{equation}
Thus, we can write Eq. (\ref{(3.31)}) as (hereafter the convention on the summation over repeated indices is employed)
\begin{equation}
\mathcal{A}_{ir}(x_j,\dot{x}_j) \ddot{x}_r + \mathcal{H}_i (x_j,\dot{x}_j)=0, \;\;\;\;\;\;\; (i=1,2).
\label{(4.3)}
\end{equation}
Bearing in mind the obvious relations 
\begin{equation}
\dfrac{{\rm d}}{{\rm d}t} \mathcal{A}_{ir}= \dot{\mathcal{A}}_{ir}= \dfrac{\partial \mathcal{A}_{ir}}
{\partial x_j} \dot{x}_j +\dfrac{\partial \mathcal{A}_{ir}}{\partial \dot{x}_j} \ddot{x}_j, 
\label{(4.4)}
\end{equation}
\begin{equation}
\dfrac{{\rm d}}{{\rm d}t} \mathcal{H}_i= \dot{\mathcal{H}}_i= \dfrac{\partial \mathcal{H}_i}
{\partial x_j} \dot{x}_j +\dfrac{\partial \mathcal{H}_i}{\partial \dot{x}_j} \ddot{x}_j,
\label{(4.5)}
\end{equation}
the first-order time derivative of (\ref{(4.3)}) gives
\begin{equation}
\mathcal{A}_{ir} \dfrac{{\rm d}^2}{{\rm d}t^2} \dot{x}_r + \left(\dfrac{\partial \mathcal{A}_{ir}}
{\partial x_j} \dot{x}_j +\dfrac{\partial \mathcal{A}_{ir}}{\partial \dot{x}_j} \ddot{x}_j \right) 
\dfrac{{\rm d}}{{\rm d}t} \dot{x}_r + \dfrac{\partial \mathcal{H}_i}{\partial x_j} \dot{x}_j 
+\dfrac{\partial \mathcal{H}_i}{\partial \dot{x}_j} \ddot{x}_j=0, \;\;\;\;\;\;\; (i=1,2),
\label{(4.6)}
\end{equation}
which represents a nonlinear system of differential equations whose solutions are given by the functions 
$\dot{x}_r$ ($r=1,2$). The second-order time derivative of (\ref{(4.3)}) leads to the system
\begin{equation}
\begin{split}
& \A \dfrac{{\rm d}^2}{{\rm d}t^2} \ddot{x}_r + 2 \left(\dfrac{\partial \A}{\partial x_j} 
\dot{x}_j +\dfrac{\partial \A}{\partial \dot{x}_j} \ddot{x}_j \right) \dfrac{{\rm d}}{{\rm d}t} \ddot{x}_r 
+ \ddot{x}_r \dfrac{\partial \A}{\partial \dot{x}_k} \dfrac{{\rm d}}{{\rm d}t} \ddot{x}_k 
+ \dfrac{\partial \H}{\partial \dot{x}_k} \dfrac{{\rm d}}{{\rm d}t} \ddot{x}_k 
+  \Biggl( \dfrac{\D^2 \A}{\D x_l \D x_k} \dot{x}_l \dot{x}_k
+ 2 \dfrac{\D^2 \A}{\D \dot{x}_l \D x_k} \ddot{x}_l \dot{x}_k \\
& + \dfrac{\D \A}{\D x_k} \ddot{x}_k + \dfrac{\D^2 \A}{\D \dot{x}_l \D \dot{x}_k} 
\ddot{x}_l \ddot{x}_k \Biggr) \ddot{x}_r+ \dfrac{\D^2 \H}{\D x_l \D x_k} \dot{x}_l \dot{x}_k
+ 2 \dfrac{\D^2 \H}{\D \dot{x}_l \D x_k} \ddot{x}_l \dot{x}_k + \dfrac{\D \H}{\D x_k} \ddot{x}_k 
+ \dfrac{\D^2 \H}{\D \dot{x}_l \D \dot{x}_k} \ddot{x}_l \ddot{x}_k=0, \;\;\;\;\;\;\; (i=1,2),
\end{split}
\label{(4.7)}
\end{equation}
still representing a nonlinear set of differential equations in the unknowns $\ddot{x}_r$. However, by 
patiently applying the time derivative operator four times to Eq. (\ref{(4.3)}) (see Appendix \ref{AppendixA} 
for details) the fourth time derivatives $x^{(4)}_r$ are found to solve a {\it linear} system of 
ordinary differential equations of the form
\begin{equation}
\begin{split}
& \A \dfrac{{\rm d}^2}{{\rm d}t^2} {x}^{(4)}_r + 4 \left(\dfrac{\partial \A}{\partial x_j} \dot{x}_j 
+\dfrac{\partial \A}{\partial \dot{x}_j} \ddot{x}_j \right) \dfrac{{\rm d}}{{\rm d}t} {x}^{(4)}_r 
+ \ddot{x}_r \left(\dfrac{\D \A}{\D \dot{x}_k}\right) \dfrac{{\rm d}}{{\rm d}t} {x}^{(4)}_k 
+ \left(\dfrac{\D \H}{\D \dot{x}_k} \right) \dfrac{{\rm d}}{{\rm d}t} {x}^{(4)}_k 
+ F_i \left(x_p,\dot{x}_p,\dots,x^{(4)}_p\right)=0, \\
&  (i,p=1,2).
\end{split}
\label{(4.8)}
\end{equation} 
The form of $F_i \left(x_p,\dot{x}_p,\dots,x^{(4)}_p\right)$ can be read from Eq. (\ref{F_i}), 
where it is explicitly shown that this term does not depend on time derivatives of $x_p$ having an order 
higher than the fourth and that its dependence on $x^{(4)}_p$ turns out to be linear. Therefore, we have 
obtained the original result according to which the differential equations describing the motion 
of a body in our Solar System within the first post-Newtonian approximation can be put in linear form if the 
fourth time derivative of the original equation is employed. In our analysis, we have been inspired 
by the work in Ref. \cite{Bruhat}, where the author differentiated (see below) a system of nonlinear 
partial differential equations to arrive at their solution. More precisely, starting from a 
system of $n$ nonlinear second-order hyperbolic partial differential equations of the form
\begin{equation}
A^{\lambda \mu} \dfrac{\D^2 w_s}{\D x^{\lambda} \D x^{\mu}} + f_s=0, \; \;\; \;\; \;\; \; 
s=1,2,\dots,n, \; \; \; \; \lambda,\mu=1,2,3,4,
\label{(4.9)}
\end{equation}  
the coefficients $A^{\lambda \mu}$ and $f_s$ depending in a nonlinear way on the four variables $x^\alpha$, 
the unknown functions $w_s$, and their first time derivatives $\dfrac{\D w_s}{\D x^\alpha}$, in Ref. \cite{Bruhat} 
it is shown that by applying five times of the derivative operator with respect to any whatsoever variable 
$x^{\alpha}$ leads to the linear system
\begin{equation}
A^{\lambda \mu} \dfrac{\D^2 U_S}{\D x^{\lambda} \D x^{\mu}}+\dfrac{\D U_S}{\D x^\lambda} + F_S=0, 
\; \;\; \;\; \;\; \; S=1,2,\dots,N, \; \; \; \; \lambda,\mu=1,2,3,4,
\footnote{$N$ represents the product by $n$ of the number of partial derivatives of order five of a function of four variables.} 
\label{(4.10)}
\end{equation}  
where $U_S$ represents the partial derivatives of order five of $w_s$ and $F_S$ is a function of the 
four variables $x^\alpha$, the unknown functions $w_s$, and of their partial derivatives up to the fifth order 
included, but not of the derivatives of higher order. In particular, when this framework is applied to Einstein's 
field equations (which represent a system of $n=10$ quasilinear partial differential equations) it suffices to 
derive them four times in order to obtain a linear system, where the unknowns are given by the fourth derivatives 
of the metric tensor components \cite{Bruhat}.  

Having obtained a linear system of differential equations of second order 
associated to (\ref{(3.31)}), we might in principle exploit their reduction to canonical form
(Appendix B) and the rich qualitative theory \cite{Poincare} of such equations. 
In fact, as we have seen in Sec. III, our model predicts the presence of periodic solutions (see for example 
Figs. \ref{Fig1.eps}, \ref{Fig2.eps}, and \ref{Fig3.eps})
in simple cases, and hence the qualitative theory just mentioned and the encouraging evidence for
simple (but nontrivial) two-body systems suggest undertaking the much harder analysis of 
$N$-body systems. 

\section{Quantum effects on Lagrangian points revisited}

The work in Secs. II and III puts on firm ground the investigations initiated in Refs.
\cite{B1,B2,B3,B4}. In other words, the motion of celestial bodies in the solar system can be studied
by employing a Lagrangian that is almost independent of the internal structure \cite{CM6}. Within
Einstein's theory, the {\it effacement} property of Newtonian theory is still valid at the first
post-Newtonian approximation, because all large, direct self-action effects cancel each other or contribute
terms in the equations of motion which can be removed, so that the final equations can be written in
terms of only some centers of mass and some effective masses \cite{CM6,CM13,CM14}. But then, to the extent
that the effective-gravity prescriptions are correct, it becomes legitimate to consider Newtonian 
potential terms among large masses in the Lagrangian used to derive geodesic motion of planets
and other bodies, and insert therein the quantum modifications worked out in Refs. \cite{EG1,EG2,EG7,EG8,EG9}. 

In particular, the work of Ref. \cite{B4} has studied in detail how quantum 
corrections on the relativistic position of the Earth-Moon Lagrangian 
points can be evaluated. In fact the effective gravity picture modifies the 
Newtonian potential among bodies of masses $m_A$ and $m_B$ in the low-energy/long-distance 
domain through the asymptotic expansion
\begin{eqnarray}
\; & \; & 
V_{E}(r) \sim -{G_N m_{A}m_{B}\over r}\left[ 1+ \left(\kappa_{1}{(l_{A}+l_{B})\over r}
+\kappa_{2}{(l_{P})^{2}\over r^{2}} +{\rm O}(G_N^{2})\right)\right]
\nonumber \\
& \Longrightarrow & {V_{E}(r) \over c^{2}m_{B}} \sim 
-{l_{A}\over r}\left[ 1+ \left(\kappa_{1}{(l_{A}+l_{B})\over r}
+\kappa_{2}{(l_{P})^{2}\over r^{2}} +{\rm O}(G_N^{2})\right)\right],
\label{asymptotic}
\label{(5.1)}
\end{eqnarray}
$l_A$ and $l_B$ being the gravitational radii of the bodies and $l_P$ the Planck length, whereas 
$\kappa_1$ and $\kappa_2$ represent constants\footnote{In this conceptual scheme, physical phenomena
are described of course in a classical way on large distances, but the precise values of some
coefficients depend on the underlying quantum theory. Thus, the dimensionless $\kappa_{1}$ parameter
is the effective-gravity weight of the purely classical term ${(l_{A}+l_{B})\over r}$.} 
resulting from the calculation of Feynman diagrams involved 
in the particular definition adopted for the potential: one-particle reducible, 
scattering or bound-states (see Tab. \ref{kappa_tab}). The term $\kappa_{1}{(l_{A}+l_{B})\over r}$ in 
(\ref{(5.1)}) refers to a post-Newtonian correction to the classical potential, whereas 
$\kappa_{2}{(l_{P})^{2}\over r^{2}}$ represents a fully quantum term, depending on the square of the Planck length. 
\begin{table}[htbp]
\centering
\caption{The values assumed by $\kappa_1$ and $\kappa_2$ in the three different potentials.}
\renewcommand\arraystretch{2.2}
\begin{tabular}{|c|c|c|c|}
\hline
$\kappa_i$ & one-particle reducible & scattering & bound-states \\
\hline
$\kappa_1$ & $-1$ & $3$ & $-\dfrac{1}{2}$ \\
\hline
$\kappa_2$ & $ -\dfrac{167}{30 \pi}$ & $\dfrac{41}{10 \pi}$ &  $\dfrac{41}{10 \pi}$ \\
\hline
\end{tabular}
\label{kappa_tab}
\end{table}

In Ref. \cite{B4} we have proposed a framework where the aforementioned quantum corrections to 
Lagrangian points can be obtained by constructing a map whose form is inspired by the pattern enlightened 
in Eq. (\ref{asymptotic}). In particular, we have applied the map 
\begin{equation}
(U_{\alpha},U_{\beta}) \rightarrow ({V}_{\alpha},{V}_{\beta}),
\label{(5.2)}
\end{equation}
with
\begin{equation}
U_{\alpha}(r) \equiv {l_{\alpha}\over r}=U_{\alpha},
\label{(5.3)}
\end{equation}
\begin{equation}
U_{\beta}(s) \equiv {l_{\beta}\over s}=U_{\beta},
\label{(5.4)}
\end{equation}
\begin{eqnarray}
{V}_{\alpha}(r) & \sim & 
\left[1+\kappa_{2}{(l_{P})^{2}\over r^{2}}\right]U_{\alpha}(r)
+\kappa_{1}(U_{\alpha}(r))^{2}
+{\rm O}(G^3_N),
\label{(5.5)}
\end{eqnarray}
\begin{eqnarray}
{V}_{\beta}(s) & \sim & \left[1+\kappa_{2}{(l_{P})^{2}\over s^{2}}\right]U_{\beta}(s)
+\kappa_{1} (U_{\beta}(s))^{2}+{\rm O}(G_N^{3}),
\label{(5.6)}
\end{eqnarray}
to the Lagrangian describing the motion of the planetoid in the gravitational field generated by the 
Earth and the Moon, which, upon adopting the set of coordinates $(ct,\xi,\eta,\zeta)$ ($c$ being the speed of light) reads as
\begin{equation}
L={1 \over 2} \sum_{\mu,\nu=0}^{3}
g_{\mu \nu}{{\rm d}x^{\mu}\over {\rm d}t}{{\rm d}x^{\nu}\over {\rm d}t},
\label{(5.7)}
\end{equation}
where 
\begin{eqnarray}
g_{00}&=& 1-2{l_{\alpha}\over r}-2{l_{\beta}\over s}-{\Omega^{2}\over c^{2}}(\xi^{2}+\eta^{2})
+2\left[\left({l_{\alpha}\over r}\right)^{2}+\left({l_{\beta}\over s}\right)^{2}\right]
\nonumber \\
&-& 2 {(l_{\alpha}+l_{\beta})\over R^{3}}
\left({l_{\alpha}\over r}+{l_{\beta}\over s}\right)(\xi^{2}+\eta^{2})
+4{l_{\alpha}\over r}{l_{\beta}\over s} 
\nonumber \\
&+& {(2-\rho)\over (1+\rho)}{l_{\alpha}\over r}{l_{\beta}\over R}
+{(2\rho-1)\over (1+\rho)}{l_{\beta}\over s}{l_{\alpha}\over R}
-7{\xi \over R^{2}}\left({l_{\alpha}\over r}l_{\beta}-{l_{\beta}\over s}l_{\alpha}\right)
\nonumber \\
&+& (1+\rho)^{-1}{\eta^{2}\over R}\left[\rho \left({l_{\alpha}\over r}\right)^{3}
{l_{\beta}\over (l_{\alpha})^{2}}
+\left({l_{\beta}\over s}\right)^{3}{l_{\alpha}\over (l_{\beta})^{2}}\right],
\label{(5.8)}
\end{eqnarray}
\begin{equation}
2c g_{01}=\left(1+2{l_{\alpha}\over r}+2{l_{\beta}\over s}\right)2 \Omega \eta,
\label{(5.9)}
\end{equation}
\begin{equation}
2c g_{02}=-\left(1+2{l_{\alpha}\over r}+2{l_{\beta}\over s}\right)2 \Omega \xi
-8{\Omega^{2} R \over (1+\rho)}\left(\rho {l_{\alpha}\over r}-{l_{\beta}\over s}\right),
\label{(5.10)}
\end{equation}
\begin{equation}
g_{03}=0,
\label{(5.11)}
\end{equation}
\begin{equation}
g_{ij}=-\left(1+2{l_{\alpha}\over r}+2{l_{\beta}\over s}\right)\delta_{ij}, \;\;\;\;  i,j=1,2,3.
\label{(5.12)}
\end{equation}
$l_\alpha$ and $l_\beta$ being the gravitational radii of the Earth and the Moon, respectively, 
$r$ and $s$ their respective distances from the planetoid, $\rho$ the ratio between the masses 
of the Earth and the Moon, $R$ their distance, and 
\begin{equation}
\Omega \equiv \omega \left[1-{3 \over 2}{(l_{\alpha}+l_{\beta})\over R}
\left(1-{1 \over 3}{\rho \over (1+\rho)^{2}}\right)\right],
\label{(5.13)}
\end{equation}
where $\omega$ represents the Newtonian pulsation.

However, since effective field theories of gravity provide quantum corrections to the Newtonian potential 
among bodies but not to its powers, we now consider a more refined prescription
where only the purely Newtonian terms (i.e., those which are linear or bilinear in $U_{\alpha}$ 
and $U_{\beta}$) are corrected through the map (5.5) and (5.6), while the remaining ones are left unchanged. 
According to this new perspective, the effective gravity Lagrangian $L_V$ 
can be obtained from (\ref{(5.7)})-(\ref{(5.12)}) by setting
\begin{equation}
\begin{split}
& U_{\alpha} \rightarrow V_\alpha, \\
& U_{\beta} \rightarrow V_\beta, \\
& U_{\alpha} U_{\beta} \rightarrow V_\alpha V_\beta, \\
& \left(U_{\alpha}\right)^n \rightarrow \left(U_\alpha\right)^n, \; \; \; \; \; n >1, \\
& \left(U_{\beta}\right)^n \rightarrow \left(U_\beta\right)^n, \; \; \; \; \; n >1.
\end{split}
\label{(5.14)}
\end{equation}
The new map (\ref{(5.14)}) is such that the quantum corrected Lagrangian reads as
\begin{eqnarray}
L_V &=&  {c^{2}\over 2} \biggr \{ 1-2({V}_{\alpha}+{V}_{\beta})
-{\Omega^{2}\over c^{2}}(\xi^{2}+\eta^{2})
+2 \left[({U}_{\alpha})^{2}+({U}_{\beta})^{2}\right] 
\nonumber \\
&-& 2{(l_{\alpha}+l_{\beta})\over R^{3}}(\xi^{2}+\eta^{2})
({V}_{\alpha}+{V}_{\beta})
+4{V}_{\alpha}{V}_{\beta} 
\nonumber \\
&+& {(2-\rho)\over (1+\rho)}{l_{\beta}\over R}{V}_{\alpha}
+{(2 \rho-1)\over (1+\rho)}{l_{\alpha}\over R}{V}_{\beta}
-7{\xi \over R^{2}}(l_{\beta}{V}_{\alpha}
-l_{\alpha}{V}_{\beta})  
\nonumber \\
&+& (1+\rho)^{-1}{\eta^{2}\over R} \left[\rho {l_{\beta}\over (l_{\alpha})^{2}}
({U}_{\alpha})^{3}
+{l_{\alpha}\over (l_{\beta})^{2}}({U}_{\beta})^{3}\right] \biggr \}  
\nonumber \\
&-& {1 \over 2}\Bigr({\dot \xi}^{2}+{\dot \eta}^{2}+{\dot \zeta}^{2}\Bigr)
\Bigr[1+2({V}_{\alpha}+{V}_{\beta})\Bigr]
+\Omega \eta {\dot \xi}\Bigr[1+2({V}_{\alpha}+{V}_{\beta})\Bigr] 
\nonumber \\
&-& \Omega \xi {\dot \eta} \Bigr[1+2({V}_{\alpha}+{V}_{\beta})\Bigr]
-4{\Omega^{2}R \over (1+\rho)}{\dot \eta}(\rho {V}_{\alpha}-{V}_{\beta}).
\label{(5.15)}
\end{eqnarray}
Therefore, new values of quantum corrections on the relativistic distances in the Earth-Moon system \cite{B4}
\begin{equation}
\begin{split}
& r_{1,GR}= 3.2637628817407598555 \times 10^8 \;  {\rm m},\\
& r_{2,GR}=  4.4892056003414800050  \times 10^8 \;  {\rm m},\\
& r_{3,GR}= 3.8167471569392170594 \times 10^8 \;  {\rm m}, \\
& r_{4,GR}= r_{5,GR}=3.8439999999998611069 \times 10^{8} {\rm m},
\end{split}
\label{(5.16)}
\end{equation}
are found and displayed in Tab. \ref{new_quantum_corrections_tab}.
Needless to say, the theoretical expression and value of such quantum corrections 
(if they exist) to Lagrangian points remain an open problem, because the classical effacement
property holds only approximately \cite{CM6,CM13,CM14}, while $3$ sets of quantum parameters
$\kappa_{1}$ and $\kappa_{2}$ are conceivable in effective gravity \cite{EG1,EG2,EG7,EG8,EG9}. 
In this respect, the quantum corrections to general relativity values for noncollinear points
$L_{4}$ and $L_{5}$ look encouraging in the case of $(\kappa_{1},\kappa_{2})$ values appropriate
to scattering potential, because corrections just below a centimeter are comparable with the purely
instrumental, time-of-flight uncertainty of the geodesic positioning techniques based laser-ranging 
\cite{LR1,LR2,LR3,LR4,LR5,LR6,LR7,LR8,LR9,LR10,LR11,LR12,LR13}. However, the total error budget 
of satellite/lunar laser ranging \cite{LR1,LR2,LR3,LR4,LR5,LR6,LR7,LR8,LR9,LR10,LR11,LR12,LR13}
varies with the specific application and/or orbit, at the level of millimeter to centimeter. 

\begin{table}
\centering
\caption{Quantum corrections on the relativistic position of Earth-Moon Lagrangian points for three 
different types of potential obtained by considering the Lagrangian function (\ref{(5.15)}).}
{
\renewcommand\arraystretch{2.0}
\begin{tabular}{|c|c|c|c|}
\hline
\multicolumn{4}{|c|}{Quantum corrections on Lagrangian points}\\
\hline
\; \;  $L_i$ \; \;  &  One-particle reducible &  Scattering & Bound-states \\
\cline{1-4} 
$L_1$ & $r_{Q}-r_{GR}=-1.27 \; {\rm mm}$ & $r_{Q}-r_{GR}=3.67 \; {\rm mm}$ & $r_{Q}-r_{GR}= -0.65 \; {\rm mm}$  \\
\hline
$L_2$ & $r_{Q}-r_{GR}=-0.75 \; {\rm mm}$ & $r_{Q}-r_{GR}=2.39 \; {\rm mm}$ & $r_{Q}-r_{GR}= -0.35 \; {\rm mm}$  \\
\hline
$L_3$ & $r_{Q}-r_{GR}= -2.96 \; {\rm mm}$ & $r_{Q}-r_{GR}= 8.89 \; {\rm mm}$ & $r_{Q}-r_{GR}=-1.48 \; {\rm mm}$  \\
\hline
$L_4$ & $r_{Q}-r_{GR}= -2.98\; {\rm mm}$ & $r_{Q}-r_{GR}=8.85 \; {\rm mm}$ & $r_{Q}-r_{GR}= -1.50\; {\rm mm}$  \\
\hline
$L_5$ & $r_{Q}-r_{GR}=-2.98 \; {\rm mm}$ & $r_{Q}-r_{GR}=8.85 \; {\rm mm}$ & $r_{Q}-r_{GR}=-1.50 \; {\rm mm}$  \\
\hline
\end{tabular}
\label{new_quantum_corrections_tab}
}
\end{table}

\section{Concluding remarks and open problems}

In the first part of our paper, to prepare the ground for future work, we have provided an original synthesis
of the Levi-Civita analysis of the problem of motion of $N$ bodies in general relativity, a very difficult
problem that was studied, among the others, by Einstein himself with Infeld and Hoffman \cite{CM4}, Fock
\cite{CM5}, Levi-Civita \cite{CM6}, Damour, Soffel and Xu \cite{CM16,CM17,CM18,CM19}.

The Sun-Earth, Sun-Earth-Moon and $N$-body dynamics have been investigated in Sec. III, while Sec. IV
contains our original proof that the nonlinear ordinary differential equations for planetary motions
can be always mapped into an exact, linear system of ordinary differential equations, where the unknowns
are the fourth-order time derivatives of the original unknown functions. In Sec. V, the evaluation
of quantum corrections to location of collinear and noncollinear Lagrangian points for the planar restricted
three-body problem has been revisited, and a new set of theoretical values of such corrections for the 
Earth-Moon-planetoid system has been displayed.
It is clear from Tab. \ref{new_quantum_corrections_tab} that the few millimeters quantum corrections 
regarding the relativistic position of Lagrangian points represent a huge obstacle for future experimental 
measurements. Nevertheless, Einstein theory produces more pronounced (classical) effects on larger  
distances than those involved in the Earth-Moon system, but at the cost of increasing the efforts for reaching 
more distant planets. As an example, by applying the framework based on the Lagrangian 
function (\ref{(5.7)})-(\ref{(5.12)}) 
(see Ref. \cite{B4} for further details) to the Sun-Earth and Sun-Jupiter systems we obtain the corrections 
reported in Tabs. \ref{Sun-Earth_GR-Newton} and \ref{Sun-Jupiter_GR-Newton} \cite{Battista-PhD}.

Notably, the values of Tab. \ref{Sun-Earth_GR-Newton} agree with those of Ref. \cite{CM21}, and those of 
Tab. \ref{Sun-Jupiter_GR-Newton} with the ones reported in Refs. \cite{CM32,CM36}. Moreover, we are aware 
of the fact that many satellites are currently situated near the Sun, but unluckily none of them is planned to 
approach the Lagrangian point $L_1$ in order to test our theoretical model. Eventually, the situation becomes 
far more complicated for the Sun-Jupiter system because of the large distances involved.

In the years to come, we hope that our result in Sec. IV, jointly with the qualitative methods 
\cite{Poincare} for linear second-order ordinary 
differential equations in canonical form (see (B4) and (B5)), will lead to improved theoretical
calculations of planetary motions in the solar system, with a wide range of applications in
fundamental and applied science.

Last but not least, it will be also very interesting to compare planetary motions according to
Ref. \cite{CM5}, where the energy-momentum tensor inside and outside
bodies is obtained, with the planetary motions according to Ref. \cite{CM6}, where
the effacement property plays instead a key role as we have seen in our Sec. II.
None of the investigations of $N$-body dynamics published so far in the literature can indeed claim complete 
superiority over the others. 

\begin{table}
\centering
\caption{General relativity corrections on the position of Newtonian Lagrangian points for the Sun-Earth system 
obtained by considering the general relativity Lagrangian (\ref{(5.7)}-\ref{(5.12)}). 
The differences involved refer to the distances of the Sun from the planetoid.}
\renewcommand\arraystretch{2.0}
\begin{tabular}{|c|c|}
\hline
\multicolumn{2}{|c|}{General relativity corrections on the Sun-Earth system}\\
\hline
\; \;  $L_i$ \; \;  &  Corrections \\
\cline{1-2}
$ L_1$ & $r_{GR}-r_{cl}=  4.8 \; {\rm m}$ \\ 
\hline
$ L_2$ & $r_{GR}-r_{cl}=  -5.0  \; {\rm m}$ \\ 
\hline
$ L_3$ & $r_{GR}-r_{cl}=  -0.3 \;  {\rm cm}$ \\ 
\hline
$ L_{4,5}$ & $r_{GR}-r_{cl}=-0.3  \;  {\rm cm}$ \\ 
\hline
\end{tabular} 
\label{Sun-Earth_GR-Newton}
\end{table} 
\begin{table}
\centering
\caption{General relativity corrections on the position of Newtonian Lagrangian points for the Sun-Jupiter system 
obtained by employing the general relativity Lagrangian (\ref{(5.7)}-\ref{(5.12)}). 
The differences involved refer to the distances of the Sun from the planetoid.}
\renewcommand\arraystretch{2.0}
\begin{tabular}{|c|c|}
\hline
\multicolumn{2}{|c|}{General relativity corrections on the Sun-Jupiter system}\\
\hline
\; \;  $L_i$ \; \;  &  Corrections \\
\cline{1-2}
$ L_1$ & $r_{GR}-r_{cl}=  30 \; {\rm m}$ \\ 
\hline
$ L_2$ & $r_{GR}-r_{cl}=   -38 \; {\rm m}$ \\ 
\hline
$ L_3$ & $r_{GR}-r_{cl}=   -1\;  {\rm m}$ \\ 
\hline
$ L_{4,5}$ & $r_{GR}-r_{cl}= -1 \;  {\rm m}$ \\ 
\hline
\end{tabular} 
\label{Sun-Jupiter_GR-Newton}
\end{table}
 
\newpage

\begin{appendix}

\section{Linear differential equations associated to nonlinear ones} \label{AppendixA}

In this appendix we provide the details of the calculations leading to (\ref{(4.8)}) starting from 
Eq (\ref{(3.31)}) written in the form given by (\ref{(4.3)}). As explained in Sec. IV, bearing in 
mind Eqs. (\ref{(4.4)}) and (\ref{(4.5)}), the first- and second-order time derivatives of 
(\ref{(4.3)}) are given by Eqs. (\ref{(4.6)}) and (\ref{(4.7)}), respectively.

The third time derivative of (\ref{(4.3)}) gives a nonlinear system of differential equations with unknowns 
$x^{(3)}_p$ ($p=1,2$) having the form
\begin{equation}
\begin{split}
& \A \dfrac{{\rm d}^2}{{\rm d}t^2} x^{(3)}_r + 3 \left(\dfrac{\partial \A}{\partial x_j} \dot{x}_j 
+\dfrac{\partial \A}{\partial \dot{x}_j} \ddot{x}_j \right) \dfrac{{\rm d}}{{\rm d}t} x^{(3)}_r 
+ \ddot{x}_r \dfrac{\partial \A}{\partial \dot{x}_k} \dfrac{{\rm d}}{{\rm d}t} x^{(3)}_k 
+ \dfrac{\partial \H}{\partial \dot{x}_k} \dfrac{{\rm d}}{{\rm d}t} x^{(3)}_k 
+ 2 \Biggl[ \left(\dfrac{{\rm d}}{{\rm d}t}\dfrac{\D \A}{\D x_j} \right) \dot{x}_j 
+ \dfrac{\D \A}{\D x_j} \ddot{x}_j \\
& + \left(\dfrac{{\rm d}}{{\rm d}t}\dfrac{\D \A}{\D \dot{x}_j} \right) \ddot{x}_j 
+\left(\dfrac{\D \A}{\D \dot{x}_j} \right) x^{(3)}_j \Biggr] x^{(3)}_r + x^{(3)}_r\dfrac{\D \A}
{\D \dot{x}_k} x^{(3)}_k + \ddot{x}_r \left( \dfrac{{\rm d}}{{\rm d}t} \dfrac{\D \A}{\D \dot{x}_k} 
\right) x^{(3)}_k + \left( \dfrac{{\rm d}}{{\rm d}t} \dfrac{\D \H}{\D \dot{x}_k} \right) x^{(3)}_k 
+ \Biggl[ \left( \dd \dfrac{\D^2 \A}{\D x_l \D x_k} \right) \dot{x}_l \dot{x}_k \\
& + \left( \dfrac{\D^2 \A}{\D x_l \D x_k} \right) \left( \ddot{x}_l \dot{x}_k + \dot{x}_l \ddot{x}_k \right) 
+ 2 \left( \dd \dfrac{\D^2 \A}{\D \dot{x}_l \D x_k} \right) \ddot{x}_l \dot{x}_k  
+ 2 \left( \dfrac{\D^2 \A}{\D \dot{x}_l \D x_k} \right) \left(x^{(3)}_l \dot{x}_k 
+ \ddot{x}_l \ddot{x}_k \right) + \left( \dd \dfrac{\D \A}{\D x_k} \right) \ddot{x}_k \\
&  + \left( \dfrac{\D \A}{\D x_k} \right) x^{(3)}_k + \left( \dd \dfrac{\D^2 \A}
{\D \dot{x}_k \D \dot{x}_l} \right) \ddot{x}_l \ddot{x}_k + \left( \dfrac{\D^2 \A}
{\D \dot{x}_k \D \dot{x}_l} \right)\left(x^{(3)}_l \ddot{x}_k + \ddot{x}_l x^{(3)}_k \right) 
\Biggr] \ddot{x}_r + \Biggl ( \dfrac{\D^2 \A}{\D x_l \D x_k} \dot{x}_l \dot{x}_k 
+ 2 \dfrac{\D^2 \A}{\D \dot{x}_l \D x_k} \ddot{x}_l \dot{x}_k \\
& + \dfrac{\D \A}{\D x_k} \ddot{x}_k  + \dfrac{\D^2 \A}{\D \dot{x}_l \D \dot{x}_k} \ddot{x}_l 
\ddot{x}_k \Biggr)x^{(3)}_r + \left( \dd \dfrac{\D^2 \H}{\D x_l \D x_k} \right) \dot{x}_l \dot{x}_k  
+ \dfrac{\D^2 \H}{\D x_l \D x_k} \left( \ddot{x}_l \dot{x}_k + \dot{x}_l \ddot{x}_k \right) 
+2 \left( \dd \dfrac{\D^2 \H}{\D \dot{x}_l \D x_k} \right) \ddot{x}_l \dot{x}_k \\
& +2 \left( \dfrac{\D^2 \H}{\D \dot{x}_l \D x_k} \right) \left(x^{(3)}_l \dot{x}_k 
+ \ddot{x}_l \ddot{x}_k \right)+ \left( \dd \dfrac{\D \H}{\D x_k} \right) \ddot{x}_k 
+ \dfrac{\D \H}{\D x_k} x^{(3)}_k  + \left( \dd \dfrac{\D^2 \H}{\D \dot{x}_l \D \dot{x}_k} 
\right) \ddot{x}_k \ddot{x}_l + \left( \dfrac{\D^2 \H}{\D \dot{x}_l \D \dot{x}_k} \right) 
\left(x^{(3)}_l \ddot{x}_k + \ddot{x}_l x^{(3)}_k \right)=0,\\
&  (i=1,2).
\end{split} 
\label{3deriv}
\end{equation}
The nonlinearities occurring in (\ref{3deriv}) vanish if we compute the fourth time derivative. In fact, as 
anticipated in Sec. IV, by differentiating once again Eq. (\ref{3deriv}) we end up with a {\it linear} system of 
coupled ordinary differential equations for the unknown functions $x^{(4)}_p$ ($p=1,2$) which can be written as
\begin{equation}
\begin{split}
& \A \dfrac{{\rm d}^2}{{\rm d}t^2} {x}^{(4)}_r + 4 \left(\dfrac{\partial \A}{\partial x_j} \dot{x}_j 
+\dfrac{\partial \A}{\partial \dot{x}_j} \ddot{x}_j \right) \dfrac{{\rm d}}{{\rm d}t} {x}^{(4)}_r 
+ \ddot{x}_r \left(\dfrac{\D \A}{\D \dot{x}_k}\right) \dfrac{{\rm d}}{{\rm d}t} {x}^{(4)}_k 
+ \left(\dfrac{\D \H}{\D \dot{x}_k} \right) \dfrac{{\rm d}}{{\rm d}t} {x}^{(4)}_k 
+ F_i \left(x_p,\dot{x}_p,\dots,x^{(4)}_p\right)=0, \\
&  (i,p=1,2).
\end{split}
\label{4deriv}
\end{equation} 
where
\begin{equation}
\begin{split}
F_i \left(x_p,\dot{x}_p,\dots,x^{(4)}_p\right) &= 3 \left[\dd \left(\dfrac{\partial \A}{\partial x_j} 
\dot{x}_j +\dfrac{\partial \A}{\partial \dot{x}_j} \ddot{x}_j \right) \right]x^{(4)}_r 
+ 2 \Biggr\{ \dd \biggl[ \left(\dfrac{{\rm d}}{{\rm d}t}\dfrac{\D \A}{\D x_j} 
\right) \dot{x}_j + \dfrac{\D \A}{\D x_j} \ddot{x}_j 
 + \left(\dfrac{{\rm d}}{{\rm d}t}\dfrac{\D \A}{\D \dot{x}_j} \right) \ddot{x}_j \\
 & +\left(\dfrac{\D \A}{\D \dot{x}_j} \right) x^{(3)}_j \biggr] \Biggr\} x^{(3)}_r + 2 \Biggl[ 
\left(\dfrac{{\rm d}}{{\rm d}t}\dfrac{\D \A}{\D x_j} \right) \dot{x}_j + \dfrac{\D \A}{\D x_j} \ddot{x}_j  
+ \left(\dfrac{{\rm d}}{{\rm d}t}\dfrac{\D \A}{\D \dot{x}_j} \right) \ddot{x}_j +\left(\dfrac{\D \A}
{\D \dot{x}_j} \right) x^{(3)}_j \Biggr]  x^{(4)}_r \\
 &+\left(\DD \dfrac{\D \A}{\D \dot{x}_k}\right) \ddot{x}_r x^{(3)}_k + 2 \left( \dd \dfrac{\D \A }
{\D \dot{x}_k} \right)\left(  x^{(3)}_r x^{(3)}_k +\ddot{x}_r x^{(4)}_k \right)+ \left( \dfrac{\D \A }
{\D \dot{x}_k} \right) \left(x^{(4)}_r x^{(3)}_k + 2 \, x^{(3)}_r x^{(4)}_k \right) \\
 &+ \left( \DD \dfrac{\D \H}{\D \dot{x}_k} \right) x^{(3)}_k + 2 \left( \dd \dfrac{\D \H}{\D \dot{x}_k} 
\right) x^{(4)}_k  +\Biggl\{ \dd\biggl[ \left( \dd \dfrac{\D^2 \A}{\D x_l \D x_k} \right) \dot{x}_l \dot{x}_k 
+ \left( \dfrac{\D^2 \A}{\D x_l \D x_k} \right) \left( \ddot{x}_l \dot{x}_k + \dot{x}_l \ddot{x}_k \right)\\
 &  + 2 \left( \dd \dfrac{\D^2 \A}{\D \dot{x}_l \D x_k} \right) \ddot{x}_l \dot{x}_k  
+ 2 \left( \dfrac{\D^2 \A}{\D \dot{x}_l \D x_k} \right) \left(x^{(3)}_l \dot{x}_k 
+ \ddot{x}_l \ddot{x}_k \right) + \left( \dd \dfrac{\D \A}{\D x_k} \right) \ddot{x}_k   
+ \left( \dfrac{\D \A}{\D x_k} \right) x^{(3)}_k \\
 &  + \left( \dd \dfrac{\D^2 \A}{\D \dot{x}_k \D \dot{x}_l} \right) \ddot{x}_l \ddot{x}_k 
+ \left( \dfrac{\D^2 \A}{\D \dot{x}_k \D \dot{x}_l} \right)\left(x^{(3)}_l \ddot{x}_k 
+ \ddot{x}_l x^{(3)}_k \right) \biggr] \Biggr \} \ddot{x}_r + \Biggl [ \left( \dd \dfrac{\D^2 \A}
{\D x_l \D x_k} \right) \dot{x}_l \dot{x}_k \\
& + \left( \dfrac{\D^2 \A}{\D x_l \D x_k} \right) \left( \ddot{x}_l \dot{x}_k + \dot{x}_l 
\ddot{x}_k \right) + 2 \left( \dd \dfrac{\D^2 \A}{\D \dot{x}_l \D x_k} \right) \ddot{x}_l \dot{x}_k  
+ 2 \left( \dfrac{\D^2 \A}{\D \dot{x}_l \D x_k} \right) \left(x^{(3)}_l \dot{x}_k 
+ \ddot{x}_l \ddot{x}_k \right) \\
&  + \left( \dd \dfrac{\D \A}{\D x_k} \right) \ddot{x}_k 
 + \left( \dfrac{\D \A}{\D x_k} \right) x^{(3)}_k + \left( \dd \dfrac{\D^2 \A}{\D \dot{x}_k 
\D \dot{x}_l} \right) \ddot{x}_l \ddot{x}_k + \left( \dfrac{\D^2 \A}{\D \dot{x}_k \D \dot{x}_l} 
\right)\left(x^{(3)}_l \ddot{x}_k + \ddot{x}_l x^{(3)}_k \right) \Biggr] x^{(3)}_r \\
& + \Biggl[ \dd \biggl ( \dfrac{\D^2 \A}{\D x_l \D x_k} \dot{x}_l \dot{x}_k + 2 \dfrac{\D^2 \A}
{\D \dot{x}_l \D x_k} \ddot{x}_l \dot{x}_k    + \dfrac{\D \A}{\D x_k} \ddot{x}_k  
+ \dfrac{\D^2 \A}{\D \dot{x}_l \D \dot{x}_k} \ddot{x}_l \ddot{x}_k \biggr) \Biggr] x^{(3)}_r 
+ \Biggl ( \dfrac{\D^2 \A}{\D x_l \D x_k} \dot{x}_l \dot{x}_k \\
& + 2 \dfrac{\D^2 \A}{\D \dot{x}_l \D x_k} \ddot{x}_l \dot{x}_k  + \dfrac{\D \A}{\D x_k} \ddot{x}_k  
+ \dfrac{\D^2 \A}{\D \dot{x}_l \D \dot{x}_k} \ddot{x}_l \ddot{x}_k  \Biggr) x^{(4)}_r 
+ \left( \DD \dfrac{\D^2 \H}{\D x_l \D x_k} \right) \dot{x}_l \dot{x}_k \\
& +2 \left( \dd \dfrac{\D^2 \H}{\D x_l \D x_k} \right)\left( \ddot{x}_l \dot{x}_k 
+ \dot{x}_l \ddot{x}_k \right) + \left(\dfrac{\D^2 \H}{\D x_l \D x_k} \right) \left( x^{(3)}_l \dot{x}_k +2 \, 
\ddot{x}_l \ddot{x}_k + \dot{x}_l x^{(3)}_k \right) + 2\left(\DD \dfrac{\D^2 \H}
{\D \dot{x}_l \D x_k} \right) \ddot{x}_l \dot{x}_k \\
& + 4 \left(\dd \dfrac{\D^2 \H}{\D \dot{x}_l \D x_k} \right) \left(x^{(3)}_l \dot{x}_k 
+ \ddot{x}_l \ddot{x}_k \right) + 2\left( \dfrac{\D^2 \H}{\D \dot{x}_l \D x_k} \right) 
\left(x^{(4)}_l \dot{x}_k  +2\, x^{(3)}_l \ddot{x}_k+\ddot{x}_l x^{(3)}_k\right) 
+\left(\DD \dfrac{\D \H}{\D x_k} \right) \ddot{x}_k \\
& + 2 \left(\dd \dfrac{\D \H}{\D x_k} \right)x^{(3)}_k + \dfrac{\D \H}{\D x_k} x^{(4)}_k 
+ \left( \DD \dfrac{\D^2 \H}{\D \dot{x}_l \D \dot{x}_k} \right) \ddot{x}_l \ddot{x}_k 
+ 2 \left( \dd \dfrac{\D^2 \H}{\D \dot{x}_l \D \dot{x}_k} \right) \left( x^{(3)}_l \ddot{x}_k 
+ \ddot{x}_l x^{(3)}_k \right) \\
& + \dfrac{\D^2 \H}{\D \dot{x}_l \D \dot{x}_k} \left( x^{(4)}_l \ddot{x}_k +2\, x^{(3)}_k x^{(3)}_l
+ \ddot{x}_l x^{(4)}_k \right), \;\;\;\;\;\;\;\;\;\; (i,p=1,2).
\end{split}
\label{F_i}
\end{equation}
It is thus clear from last equation that the functions $F_i \left(x_p,\dot{x}_p,\dots,x^{(4)}_p\right)$ 
($i=1,2$) depend linearly on $x^{(4)}_p$ ($p=1,2$) and that no derivatives of order higher than four appear, 
which clearly means that Eq. (\ref{4deriv}) (or equivalently (\ref{(4.8)})) is linear with respect to 
$x^{(4)}_p$.\footnote{Recall that the most general form of a linear ordinary differential equation of order $n$ is given by
\begin{equation}
\left( r_n(t) \dfrac{{\rm d}^n}{{\rm d}t^{n}} + r_{n-1}(t) \dfrac{{\rm d}^{n-1}}{{\rm d}t^{n-1}} 
+ \dots + r_1(t) \dfrac{{\rm d}}{{\rm d}t} + r_0(t) \right) x(t)=f(t),
\end{equation}
the coefficients $r_n(t), r_{n-1}(t), \dots,r_1(t),r_0(t)$ and the term $f(t)$ being continuous real-valued 
functions of $t$ in the interval $a\leq t\leq b$ and $r_n(t)$ a function that does not vanish at any point of 
the aforementioned interval. The operator
\begin{equation}
L \equiv r_n(t) \dfrac{{\rm d}^n}{{\rm d}t^{n}} + r_{n-1}(t) \dfrac{{\rm d}^{n-1}}{{\rm d}t^{n-1}} 
+ \dots + r_1(t) \dfrac{{\rm d}}{{\rm d}t} + r_0(t),
\end{equation}
is called linear differential operator of order $n$.}

\section{Linear differential equations of second order. Sturm-Liouville problem}\label{Sturm-Liouville}

Within the framework of ordinary differential equations, every second-order linear differential
equation can be written in the form
\begin{equation}
\left[{d^{2}\over dx^{2}}+p(x){d \over dx}+q(x) \right]u(x)=0,
\label{(B1)}
\end{equation}
where $x$ is taken to lie in the closed interval $[a,b]$, while $p$ and $q$ are suitably
smooth functions. This equation can be brought into the Liouville form, where the coefficient
of the first-order derivative vanishes. For this purpose, one sets $u(x)=\varphi(x) \psi(x)$, so that
Eq. (\ref{(B1)}) reads as
\begin{equation}
\varphi \psi''+(2 \varphi'+p \varphi)\psi'+(\varphi''+p \varphi'+q \varphi)\psi=0.
\label{(B2)}
\end{equation}
Our task is achieved if the function $\varphi$ solves the first-order equation
\begin{equation}
{\varphi' \over \varphi}=-{p \over 2}
\Longrightarrow \varphi(x)={\rm exp}\left(-{1 \over 2}\int p(x)dx \right).
\label{(B3)}
\end{equation}
At this stage, since we can divide by $\varphi$, we can re-express Eq. (\ref{(B2)}) in the form
\begin{equation}
\psi''+W \psi=0,
\label{(B4)}
\end{equation}
where the potential terms turns out to be
\begin{equation}
W={\varphi'' \over \varphi}+p{\varphi' \over \varphi}+q
=-{1 \over 2}p'-{p^{2}\over 4}+q.
\label{(B5)}
\end{equation}
Once we have reduced ourselves to studying Eq. (\ref{(B4)}), one can deduce important qualitative properties. 
For example, if the function $W$ is continuous for $x \in [a,b]$, and if there exist real constants
$\omega$ and $\Omega$ such that
\begin{equation}
0 < \omega^{2} < W(x) < \Omega^{2},
\label{(B6)}
\end{equation}
one can compare the zeros of solutions of Eq. (\ref{(B4)}) with the zeros of solutions
of the equations
\begin{equation}
\psi''+\alpha^{2}\psi=0, \;\; \; \;\;\; \; \; \alpha=\omega \; {\rm or} \; \Omega.
\label{(B7)}
\end{equation}
Equations (\ref{(B7)}) are solved by periodic functions $\sin \alpha(x-x_{0})$ which have zeros at
$x_{0}+{k \pi \over \alpha}$, $k$ being an integer and $\alpha$ taking one of the two values in
(\ref{(B7)}). One can then prove that the difference $\delta$ between two adjacent zeros of a solution
of Eq. (\ref{(B4)}) satifies the conditions \cite{Valiron1945}
\begin{equation}
{\pi \over \Omega} \leq \delta \leq {\pi \over \omega}.
\label{(B8)}
\end{equation}

Now the first of our Eqs. (\ref{(3.15)}) (or the more general (\ref{(3.31)})) 
can be written, after using the procedure used in Sec. IV, 
i.e. four differentiations with respect to $t$ of the original equation, in the form 
\begin{equation}
g(t)\left[Y''+p_{1}(t)Y'+q_{1}Y \right]
+f(t)\left[X''+p_{2}(t)X'+q_{2}X \right]=0,
\label{(B9)}
\end{equation}
where, for each pair $(p_{1},q_{1})$ and $(p_{2},q_{2})$, we can evaluate the corresponding potential
$W_{1}$ (respectively $W_{2}$) according to Eq. (B5). The letters $X$ and $Y$ here denote fourth-order 
time derivative of the original functions $x(t)$ and $y(t)$, respectively. 

\end{appendix}

\acknowledgments
E. B. and G. E. are grateful to the Dipartimento di Fisica ``Ettore Pancini'' of Federico II University for
hospitality and support. The work of E. B., S. D. and G. E. has been supported by the 
INFN funding of the NEWREFLECTIONS experiment.


\begin{thebibliography}{99}

\bibitem{B1}
E. Battista and G. Esposito, Restricted three-body problem in effective-field-theory
models of gravity, Phys. Rev. D {\bf 89}, 084030 (2014).
\bibitem{B2}
E. Battista and G. Esposito, Full three-body problem in effective-field-theory
models of gravity, Phys. Rev. D {\bf 90}, 084010 (2014); Phys. Rev. D {\bf 93}, 049901(E) (2016).
\bibitem{B3}
E. Battista, S. Dell'Agnello, G. Esposito and J. Simo, Quantum effects on Lagrangian points and
displaced periodic orbits in the Earth-Moon system, Phys. Rev. D {\bf 91}, 084041 (2015);
Phys. Rev. D {\bf 93}, 049902(E) (2016).
\bibitem{B4}
E. Battista, S. Dell'Agnello, G. Esposito, L. Di Fiore, J. Simo and A. Grado, Earth-Moon Lagrangian 
points as a test bed for general relativity and effective field theories of gravity,
Phys. Rev. D {\bf 92}, 064045 (2015); Phys. Rev. D {\bf 93}, 109904(E) (2016).
\bibitem{EG1}
J. F. Donoghue, Leading quantum correction to the Newtonian potential, 
Phys. Rev. Lett. {\bf 72}, 2996 (1994).
\bibitem{EG2}
J. F. Donoghue, General relativity as an effective field theory: the leading 
quantum corrections, Phys. Rev. D {\bf 50}, 3874 (1994).
\bibitem{EG3}
I. J. Muzinich and S. Vokos, Long range forces in quantum gravity, Phys. Rev. D {\bf 52}, 3472 (1995).
\bibitem{EG4}
H. W. Hamber and S. Liu, On the quantum corrections to the Newtonian potential, Phys. Lett. B {\bf 357}, 51 (1995).
\bibitem{EG5}
A. A. Akhundov, S. Bellucci, and A. Shiekh, Gravitational interaction to one loop in effective 
quantum gravity, Phys. Lett. B {\bf 395}, 16 (1997).
\bibitem{EG6}
I. B. Khriplovich and G. G. Kirilin, Quantum power correction to the Newton law, Sov. Phys. JETP {\bf 95}, 981 (2002).
\bibitem{EG7}
N. E. J. Bjerrum-Bohr, J. F. Donoghue, and B. R. Holstein, Quantum gravitational
corrections to the nonrelativistic scattering potential of two masses,
Phys. Rev. D {\bf 67}, 084033 (2003).
\bibitem{EG8}
N. E. J. Bjerrum-Bohr, J. F. Donoghue, and B. R. Holstein, Quantum corrections to the
Schwarzschild and Kerr metrics, Phys. Rev. D {\bf 68}, 084005 (2003).
\bibitem{EG9}
J. F. Donoghue, The effective field theory treatment of quantum gravity, 
AIP Conf. Proc. {\bf 1483}, 73 (2012).
\bibitem{CM1}
F. F. Tisserand, {\it Treatise of Celestial Mechanics}, Vols. 1 to 4
(Gauthier-Villars, Paris, 1889-1896; Jacques Gabay, Paris, 1990).
\bibitem{CM2}
H. Poincar\'e, The three-body problem and the equations of dynamics,
Acta Mathematica {\bf 13}, 1 (1890); On the three-body problem, 
Bull. Astronomique {\bf 8}, 12 (1891).
\bibitem{CM3}
H. Poincar\'e, {\it Les Methodes Nouvelles de la Mecanique Celeste} (Gauthier-Villars, Paris, 1892),
reprinted as {\it New Methods of Celestial Mechanics}, edited by D. L. Goroff (American Institute
of Physics, College Park, 1993).
\bibitem{CM4}
A. Einstein, L. Infeld, and B. Hoffman, The gravitational equations and the problem of motion,
Ann. Math. {\bf 39}, 65 (1938).
\bibitem{CM5}
V. A. Fock, On the motion of finite masses after the Einstein theory of gravitation,
J. Phys. (Moscow) {\bf 1}, 81 (1939).
\bibitem{CM6}
T. Levi-Civita, {\it The N-Body Problem in General Relativity} (Gauthier-Villars, Paris, 1941;
Reidel, Dordrecht, 1964).
\bibitem{CM7}
B. D. Tapley and J. M. Lewallen, Solar influence on satellite motion near the stable
earth-moon libration points, AIAA J. {\bf 2}, 728 (1964).
\bibitem{CM8}
L. A. Pars, {\it A Treatise on Analytical Dynamics} (Heinemann, London, 1965).
\bibitem{CM9}
V. Szebehely, {\it Theory of Orbits: the Restricted Problem of Three Bodies} (Academic Press,
New York, 1967).
\bibitem{CM10}
E. Krefetz, Restricted three-body problem in the post-Newtonian approximation, 
Astron. J. {\bf 72}, 471 (1967).
\bibitem{CM11}
V. A. Brumberg, {\it Relativistic Celestial Mechanics} (Nauka, Moscow, 1972).
\bibitem{CM12}
R. A. Freitas and F. Valdes, A search for natural or artificial objects located at the Earth-Moon 
libration points, Icarus {\bf 42}, 442 (1980).
\bibitem{CM13}
T. Damour, The problem of motion in Newtonian and Einsteinian gravity, in
{\it 300 Years of Gravitation}, eds. S. W. Hawking and W. Israel (Cambridge University Press,
Cambridge, 1987).
\bibitem{CM14}
T. Damour and G. Schafer, Levi-Civita and the general relativistic problem of motion, 
in {\it Studies in the History of General Relativity, Einstein Studies Vol. 3},
eds. J. Eisenstaedt, A. J. Kox (Birkhauser, Boston, 1992).
\bibitem{CM15}
A. Celletti and A. Giorgilli, On the stability of the Lagrangian points in the spatial restricted
problem of three bodies, Cel. Mech. Dyn. Astron. {\bf 50}, 31 (1991).
\bibitem{CM16}
T. Damour, M. Soffel, and C. Xu, General-relativistic celestial mechanics. I. Method and
definition of reference systems, Phys. Rev. D {\bf 43}, 3273 (1991).
\bibitem{CM17}
T. Damour, M. Soffel, and C. Xu, General-relativistic celestial mechanics. II. Translational
equations of motion, Phys. Rev. D {\bf 45}, 1017 (1992).
\bibitem{CM18}
T. Damour, M. Soffel, and C. Xu, General-relativistic celestial mechanics. III. Rotational
equations of motion, Phys. Rev. D {\bf 47}, 3124 (1993).
\bibitem{CM19}
T. Damour, M. Soffel, and C. Xu, General-relativistic celestial mechanics. IV. Theory of
satellite motion, Phys. Rev. D {\bf 49}, 618 (1994).
\bibitem{CM20}
T. I. Maindl and R. Dvorak, On the dynamics of the relativistic restricted three-body problem,
Astronomy \& Astrophys. {\bf 290}, 335 (1994).
\bibitem{CM21}
T. I. Maindl, The solar system's Lagrangian points in the framework of the relativistic restricted
three-body problem, ASP Conference Series {\bf 107}, 147 (1996).
\bibitem{CM22}
A. Celletti and L. Chierchia, On the stability of realistic three-body problems, 
Commun. Math. Phys. {\bf 186}, 413 (1997).
\bibitem{CM23}
G. Dell'Antonio, Noncollision periodic solutions of the N-body system,
Nonlinear Differ. Equ. Appl. {\bf 5}, 117 (1998).
\bibitem{CM24}
K. B. Bhatnagar and P. P. Hallan, Existence and stability of $L_{4,5}$ in the relativistic restricted
three-body problem, Cel. Mech. Dyn. Astron. {\bf 69}, 271 (1998).
\bibitem{CM25}
C. D. Murray and S. F. Dermott, {\it Solar System Dynamics} (Cambridge University Press, Cambridge, 1999).
\bibitem{CM26}
C. N. Douskos and E. A. Perdios, On the stability of equilibrium points in the relativistic restricted
three-body problem, Cel. Mech. Dyn. Astron. {\bf 82}, 317 (2002).
\bibitem{CM27}
A. Morbidelli, {\it Modern Celestial Mechanics: Aspects of Solar System Dynamics} (Taylor \& Francis, London, 2002). 
\bibitem{CM28}
V. A. Brumberg, Special solutions in a simplified restricted three-body problem with gravitational
radiation taken into account, Cel. Mech. Dyn. Astron. {\bf 85}, 269 (2003).
\bibitem{CM29}
L. F. Wanex, Chaotic amplification in the relativistic restricted three-body problem,
Z. Naturforsch. {\bf 58}a, 13 (2003).
\bibitem{CM30}
S. Kopeikin and I. Vlasov, The effacing principle in the post-Newtonian celestial mechanics,
in Proceedings XI Marcel Grossmann Meeting, 2475-2477 (World Scientific, Singapore, 2008).
\bibitem{CM31}
H. Asada, Gravitational wave forms for a three-body system in Lagrange's orbit: Parameter determinations 
and a binary source test, Phys. Rev. D {\bf 80}, 064021 (2009).
\bibitem{CM32}
K. Yamada and H. Asada, Collinear solution to the general relativistic three-body problem, 
Phys. Rev. D {\bf 82}, 104019 (2010).
\bibitem{CM33}
H. Asada, T. Futamase and P. Hogan, {\it Equations of Motion in General Relativity},
Int. Ser. Monogr. Phys. {\bf 148} (Oxford University Press, Oxford, 2010).
\bibitem{CM34}
M. Connors, P. Wiegert, and C. Veillet, Earth's Trojan asteroid, Nature (London) {\bf 475}, 481 (2011).
\bibitem{CM35}
B. Bertotti, P. Farinella and D. Vokrouhlicky, {\it Physics of the Solar System: 
Dynamics and Evolution, Space Physics, and Spacetime Structure} (Springer, Berlin, 2012).
\bibitem{CM36}
K. Yamada and H. Asada, Triangular solution to the general relativistic three-body problem
for general masses, Phys. Rev. D {\bf 86}, 124029 (2012).
\bibitem{CM37}
{\it List of Jupiter Trojans}, Minor Planet Center, 25 February 2014.
\bibitem{CM38}
K. Yamada and H. Asada, Post-Newtonian effects on the stability of the triangular solution
in the three-body problem for general masses, Phys. Rev. D {\bf 91}, 124016 (2015).
\bibitem{CM39}
K. Yamada and H. Asada, Non-chaotic evolution of triangular configuration due to gravitational
radiation reaction in the three-body problem, Phys. Rev. D {\bf 93}, 084027 (2016).
\bibitem{CM40}
T. Y. Zhou, W. G. Cao, and Y. Xie, Collinear solution to the three-body problem under a 
scalar-tensor gravity, Phys. Rev. D {\bf 93}, 064065 (2016).
\bibitem{CM41}
F. L. Dubeibe, F. D. Lora-Clavijo, and G. A. Gonzalez, Post-Newtonian circular restricted
3-body problem: Schwarzschild primaries, arXiv:1605.06204 [gr-qc].
\bibitem{OR1}
C. R. McInnes, {\it Solar Sailing: Technology, Dynamics and Mission Applications}
(Springer Praxis, London, 1999).
\bibitem{OR2}
J. Simo and C. R. McInnes, Solar sail trajectories at the earth-moon Lagrange points, 
in {\it 59th International Astronomical Congress, Glasgow, Scotland}, 2008.
\bibitem{OR3}
J. Simo and C. R. McInnes, Displaced periodic orbits with low-thrust propulsion in the 
earth-moon system, in {\it 19th AAS/AIAA Space Flight Mechanics Meeting, Savannah, Georgia}, 2009.
\bibitem{OR4}
J. Simo and C. R. McInnes, Solar sail orbits at the earth-moon libration points, 
Comm. Nonlinear Sci. Numer. Simulat. {\bf 14}, 4191 (2009).
\bibitem{OR5}
J. Simo and C. R. McInnes, Asymptotic analysis of displaced lunar orbits, 
J. of Guidance, Control, and Dynamics {\bf 32}, 1666 (2009).
\bibitem{OR6}
J. Simo and C. R. McInnes, Displaced solar sail orbits: dynamics and applications, 
in {\it  20th AAS/AIAA Space Flight Mechanics Meeting, San Diego, California}, 2010.
\bibitem{OR7}
J. Simo and C. R. McInnes, Designing displaced lunar orbits using low-thrust propulsion, 
J. of Guidance, Control, and Dynamics {\bf 33}, 259 (2010).
\bibitem{OR8}
J. Simo and C. R. McInnes, Feedback stabilization of displaced periodic orbits: 
application to binary asteroids, Acta Astronautica {\bf 96}, 106 (2014).
\bibitem{LR1}
P. L. Bender et al., The lunar experiment, Science {\bf 182}, 229 (1973).
\bibitem{LR2}
I. I. Shapiro, R. D. Reasenberg, J. F. Chandler, and R. W. Babcock, Measurement of the de Sitter precession of 
the moon: A relativistic three-body effect, Phys. Rev. Lett. {\bf 61}, 2643 (1988).
\bibitem{LR3}
M. R. Pearlman, J. J. Degnan, and J. M. Bosworth, The international laser ranging service, Adv. Space Res. {\bf 30}, 135 (2002).
\bibitem{LR4}
J. G. Williams, S. G. Turyshev, and D. H. Boggs, Progress in lunar laser ranging tests of relativistic 
gravity, Phys. Rev. Lett. {\bf 93}, 261101 (2004).
\bibitem{LR5}
Z. Altamimi, X. Collilieux, J. Legrand et al., ITRF2005: A new release of the international terrestrial reference frame 
based on time series of station positions and earth orientation parameters, J. Geophys. Res. {\bf 112}, B09401 (2007).
\bibitem{LR6}
R. March, G. Bellettini, R. Tauraso, and S. Dell'Agnello, Constraining spacetime torsion with the Moon and Mercury, 
Phys. Rev. D {\bf 83}, 104008 (2011).
\bibitem{LR7}
R. March, G. Bellettini, R. Tauraso, and S. Dell'Agnello, Constraining spacetime torsion with LAGEOS, 
Gen. Relativ. Gravit. {\bf 43}, 3099 (2011).
\bibitem{LR8}
S. Dell'Agnello et al., Creation of the new industry-standard space test of laser retroreflectors for the 
GNSS and LAGEOS, J. Adv. Space Res. {\bf 47}, 822 (2011).
\bibitem{LR9}
S. Dell'Agnello et al., Fundamental physics and absolute positioning metrology with the MAGIA 
lunar orbiter, Exp. Astron. {\bf 32}, 19 (2011).
\bibitem{LR10}
S. Dell'Agnello et al., Probing general relativity and new physics with lunar laser ranging, 
Nucl. Instr. Methods Phys. Res. A {\bf 692}, 275 (2012).
\bibitem{LR11}
M. Martini, S. Dell'Agnello et al., MoonLIGHT: A USA-Italy lunar laser ranging retroreflector array for the 
21st century, Planet. Space Sci. {\bf 74}, 276 (2012).
\bibitem{LR12}
D. Currie, S. Dell'Agnello, G. O. Delle Monache, B. Behr, and J. G. Williams, A lunar laser ranging retroreflector 
array for the 21st century, Nucl. Phys. B Proc. Suppl. {\bf 243}, 218 (2013).
\bibitem{LR13}
S. Dell'Agnello et al., Next-generation laser retroreflectors for GNSS, solar system exploration, geodesy, 
gravitational physics and earth-observation, in ESA Proc. Int. Conf. on Space Optics (Tenerife, Spain, Oct. 2014).
\bibitem{SUN}
D. Vokrouhlicky, A note on the solar radiation perturbations of lunar motion, Icarus {\bf 126}, 293 (1997).
\bibitem{ADC}
T. Levi Civita, {\it The Absolute Differential Calculus} (Blackie \& Son, London, 1926;
Dover, New York, 1977).
\bibitem{MTW}
C. W. Misner, K. S. Thorne, and J. A. Wheeler, {\it Gravitation}, (W. H. Freeman and Company, San Francisco, 1973)
\bibitem{Battista-PhD}
E. Battista, Extreme regimes in quantum gravity, PhD thesis (Naples University, 2016); arXiv:1606.04259. 
\bibitem{Damour1}
A. Buonanno and T. Damour, Effective one-body approach to general relativistic two-body dynamics, 
Phys. Rev. D {\bf 59}, 084006 (1999).
\bibitem{Damour2}
A. Buonanno and T. Damour, Transition from inspiral to plunge in binary black hole coalescences, 
Phys. Rev. D {\bf 62}, 064015 (2000).
\bibitem{Damour3}
A. Buonanno, Y. Chen, and T. Damour, Transition from inspiral to plunge in precessing binaries of 
spinning black holes, Phys. Rev. D {\bf 74}, 104005 (2006).
\bibitem{Damour4}
T. Damour, A. Nagar, M. Hannam, S. Husa, and B. Bruegmann, Accurate effective-one-body waveforms of 
inspiralling and coalescing black-hole binaries, Phys. Rev. D {\bf 78}, 044039 (2008). 
\bibitem{Damour5}
T. Damour and A. Nagar, An improved analytical description of inspiralling and coalescing black-hole binaries, arXiv:0902.0136.
\bibitem{Damour6}
T. Damour, B. R. Iyer, and A. Nagar, Improved resummation of post-Newtonian multipolar waveforms from circularized compact binaries,
Phys. Rev. D {\bf 79}, 064004 (2009).
\bibitem{Damour7}
D. Bini, T. Damour, and A. Geralico, Confirming and improving post-Newtonian and effective-one-body results from 
self-force computations along eccentric orbits around a Schwarzschild black hole, Phys. Rev. D {\bf 93}, 064023 (2016).
\bibitem{Damour8}
D. Bini, T. Damour, and A. Geralico, New gravitational self-force analytical results for eccentric orbits 
around a Schwarzschild black hole, Phys. Rev. D {\bf 93}, 104017 (2016).
\bibitem{Damour9}
T. Damour and D. Bini, Conservative second-order gravitational self-force on circular orbits 
and the effective one-body formalism, Phys. Rev. D {\bf 93}, 104040 (2016).
\bibitem{Damour10}
D. Bini, T. Damour, and A. Geralico, High post-Newtonian order gravitational self-force analytical 
results for eccentric equatorial orbits around a Kerr black hole, Phys. Rev. D {\bf 93}, 124058 (2016).
\bibitem{Bruhat}
Y. Four\`es-Bruhat, Th\'eor\`eme d'existence pour certains syst\`emes d'\'equations aux d\'eriv\'ees 
partielles non lin\'eaires, Acta Mathematica {\bf 88}, 141 (1952); English translation in the Max Planck 
Institute for History of Science preprint series, document 480 (2016).
\bibitem{Poincare}
H. Poincar\'e, Collected Works, Vol. 1 (Gauthier-Villars, Paris, 1928; Jacques Gabais, Paris, 2011).
\bibitem{Valiron1945}
G. Valiron, {\it Functional Equations. Applications} (Masson, Paris, 1945).

\end{thebibliography}
\end{document}